    \newcommand{\be}[0]{\begin{equation}}
    \newcommand{\ee}[0]{\end{equation}}
    \newcommand{\ba}[0]{\begin{eqnarray}}
    \newcommand{\ea}[0]{\end{eqnarray}}    
        
\documentclass[12pt]{article}
\usepackage{epsfig}
\usepackage{amsmath}
\begin{document}
\def\lmsb{\ensuremath{\Lambda_{\overline{{\mathrm{MS}}}}}\,}
\def\msb{\ensuremath{\overline{{\mathrm{MS}}}}}
\def\as{\ensuremath{\alpha_s}}
\def\epem{\ensuremath{\mbox{e}^{+}\mbox{e}^{-}\,\!}}
\def\mR{\ensuremath{\mathcal{R}}} \def\mG{\ensuremath{\mathcal{G}}}
\def\rd{\mathrm{d}} 

\large
\hfill\vbox{\hbox{\tt hep-ph/0011203}\hbox{LUTP/00/48}
  \hbox{IPPP/00/06}\hbox{DTP/00/68}
  \hbox{November 2000}} 
\large 
\nopagebreak

\vspace{0.75cm}
\begin{center}
  {\LARGE {\bf Direct Extraction of QCD
      $\boldsymbol{\Lambda_{\mathrm{\scriptscriptstyle{\overline{MS}}}}}$
      from \epem\ Jet Observables}}
  \vspace{0.6cm}
  
  {\Large S.~J.~Burby\footnote{e-mail: {\tt Stephen@thep.lu.se}}}\\
  {\it Department of Theoretical Physics, Lund University,\\
    S\"olvegatan 14A, S-223 62
    Lund, Sweden}\\
  \vskip .5cm
  
  {\Large C.~J.~Maxwell\footnote{e-mail: {\tt C.J.Maxwell@durham.ac.uk}}}\\
  {\it Centre for Particle Theory,
    University of Durham,\\
    South Road, Durham DH1 3LE, England }


\end{center}

\normalsize
\vspace{0.45cm}

\centerline{\bf Abstract}
\vspace{0.3cm}
We directly fit the QCD dimensional transmutation parameter, \lmsb, to
experimental data on \epem\ jet observables, making use of next-to-leading
order (NLO) perturbative calculations. In this procedure there is no need to
mention, let alone to arbitrarily vary, the unphysical renormalisation scale
${\mu}$, and one avoids the spurious and meaningless ``theoretical error''
associated with standard \as\ determinations.  PETRA, SLD, and LEP data are
considered in the analysis. An attempt is made to estimate the importance of
uncalculated next-NLO and higher order perturbative corrections, and power
corrections, by studying the scatter in the values of \lmsb\ obtained for
different observables.

\newpage

\section*{1 Introduction}

The consistent extraction of the QCD coupling strength \as\ from experimental
data on a wide range of strong interaction processes has provided compelling
evidence for Quantum Chromodynamics as the underlying theory of this sector
of the Standard Model, for a recent comprehensive review see \cite{r1}.
Notwithstanding attempts to exploit Lattice Gauge Theory calculations [2-4],
the bulk of these determinations extract \as\ by fitting the measured data to
fixed-order perturbative predictions supplemented by computer-modelled
hadronisation corrections. The asymptotic freedom of QCD leads one to hope
that at sufficiently high energy scales (small enough values of the coupling)
these determinations will be reasonably accurate. In this paper we shall be
concerned with jet observables such as jet rates, thrust distributions,...,
studied in \epem\ collisions. For these observables next-to-leading order
(NLO) perturbative corrections have been calculated [5-8], and
non-perturbative hadronisation (power) corrections can be modelled using a
variety of computer Monte Carlo techniques
\cite{Sjostrand:1994yb,Marchesini:1992ch}. A difficulty with fixed-order
renormalisation group (RG)-improved perturbation theory is that NLO
predictions depend on the dimensionful renormalisation scale $\mu$ which
arises in removing ultraviolet divergences from the calculation. Usually one
chooses $\mu$ proportional to $Q$ the physical energy scale of the process,
for instance the c.m. energy in \epem\ collisions, so that ${\mu}=xQ$, with
$x$ an unspecified dimensionless constant. This is then varied over some
range around ${\mu}=Q$, $x=1$, the so-called ``physical scale'', say between
${\mu}=10Q$ and ${\mu}=0.1Q$. The resulting fits to data of
${\alpha}_{s}(xQ)$ are then customarily converted to ${\alpha}_{s}(M_Z)$
using the two-loop evolution of the coupling to yield a central value based
on $x=1$, and a ``theoretical error bar'' based on the variation between
$x=10$ and $x=0.1$. Unfortunately, the range of $x$ to be considered is
completely {\it ad hoc}, and there is no reason why the central values
obtained should reflect the actual value of ${\alpha}_{s}(M_Z)$.  According
to \cite{r1} this arbitrariness leads to ``infinite discomfort in finite
order''. The cure for this discomfort is to recognise that the dimensionful
scale $\mu$ has no physical relevance whatsoever.
The fact that fixed-order RG-improved perturbative predictions
depend on it is a result of the standard way in which RG-improvement is
carried out \cite{r11,r12}, with the perturbation series being truncated at
NLO, and a choice of scale which is $Q$-dependent. One should rather keep
$\mu$ {\it independent} of $Q$.  The renormalisation scheme-dependent
coupling ${\alpha}_{s}({\mu})$ then has no dependence on $Q$, the
$Q$-dependence of the observable coming entirely from the perturbative
coefficients which contain ``unphysical'' $\mu$-dependent logarithms and
``physical'' $Q$-dependent ultraviolet (UV) logarithms. Fixed-order
truncation of the perturbation series is no longer adequate, in particular
such truncations with $\mu$ independent of $Q$ do not satisfy asymptotic
freedom. The RG self-consistency of perturbation theory allows one to
identify an infinite subset of RG-predictable logarithms at any given order
of Feynman diagram calculation. If these are resummed to all-orders the
``unphysical'' $\mu$-dependent logarithms contained in ${\alpha}_{s}(\mu)$
cancel against those contained in the perturbative coefficients, and one
obtains a $\mu$-independent result, which correctly builds the leading
$Q$-dependence of the observable. This procedure has been termed ``Complete
Renormalisation Group Improvement'' (CORGI) \cite{r11,r12}, and it has been
shown how to extend the argument to problems involving factorisation scales
in addition to renormalisation scales \cite{r12}.  At NLO this approach
yields exactly the same result as the Effective Charge approach of Grunberg
\cite{r13,r14}, which corresponds to choosing $\mu$ with standard
RG-improvement in such a way that {\it all} the ``physical'' UV logarithms of
$Q$ are resummed. In general with the standard RG-improvement an infinite
subset of these logarithms is omitted, with the result that one obtains
${\mu}$-dependent results, and does not reproduce the correct physical
leading $Q$-dependence of the observable.\\

In fact it is possible to motivate this approach more straightforwardly by
showing how QCD observables may be directly related to the dimensional
transmutation parameter of the theory. In Section 2 we shall show how the
dimensional transmutation parameter arises on the grounds of generalised
dimensional analysis, modifying the analysis of \cite{r15}. The form of this
relation at the NLO level is then completely equivalent to the CORGI approach
outlined above. The advantage of this derivation is that mention of the
renormalisation scale $\mu$ and the renormalised coupling
${\alpha}_{s}({\mu})$ can be essentially avoided, and the physical
irrelevance of these quantities is manifest. In contrast the fundamental
importance of the parameter $\Lambda$ is stressed.  In Section 3 we shall
define the QCD \epem\ jet observables with which we shall be
concerned. Section 4 contains direct plots of \lmsb\
obtained bin-by-bin in jet resolution criterion, thrust,...,etc., from
hadronisation corrected data, using the direct relation
between \lmsb\ and the data.  We attempt to estimate
the uncertainty in \lmsb\ due to uncalculated NNLO and
higher-order corrections, and possible power corrections, by looking at the
scatter in the extracted values between different observables. In Section 5
we study the dependence of the thrust distribution on the c.m. energy $Q$
using data spanning the PETRA-LEP 1-LEP 2 energy range, and we perform joint
fits for NNLO perturbative corrections and power corrections. Section 6
contains a discussion and our Conclusions.\\

\section*{2 Direct Relation between $\boldsymbol{\Lambda}$ and QCD Observables} 
In this section we shall derive from basic considerations of generalised
dimensional analysis how the dimensional transmutation parameter ${\Lambda}$
arises, and how it may be directly related to the QCD observable. The
derivation owes much to the discussion of \cite{r15} and closely follows that
of \cite{r18}. We suppose that we have a dimensionless generic QCD observable
${\mR}(Q)$, dependent on the single dimensionful (energy) scale $Q$.
Quark masses will be taken to be zero throughout our discussion, the
extension to the massive case has been considered in \cite{r19}. Since
${\mR}(Q)$ is dimensionless, dimensional analysis clearly demands that
\begin{equation} 
{\mR}(Q)={\Phi}\left(\frac{\Lambda}{Q}\right)\;, 
\end{equation}
where $\Lambda$ is a dimensionful scale , which will turn out to be related to the
dimensional transmutation parameter. There is an extra trivial possibility
that ${\mR}(Q)=C$, where $C$ is a dimensionless {\it constant}. That is,
there is no energy dependence. This trivial $Q$-dependence would be the case
if the bare coupling of QCD was finite, since the QCD Lagrangian (with
massless quarks) contains no massive parameters. Of course, in fact, the bare
coupling is infinite, and an infinite renormalisation must be performed,
leading to a functional relation as in Eq.(1). An obvious proposal is to
invert Eq.(1) to obtain 
\begin{equation} 
\frac{\Lambda}{Q}={\Phi}^{-1}({\mR}(Q)), 
\end{equation} 
where ${\Phi}^{-1}$ is the inverse function. This is indeed the basic
motivation for Grunberg's method of Effective Charges \cite{r14}. We shall
obtain the form of ${\Phi}^{-1}$ by starting from the form of the derivative
of ${\mR}(Q)$ with respect to $Q$, imposed by dimensional analysis. We
must clearly have, 
\begin{equation}
 \frac{\rd{\mR}(Q)}{\rd Q}=\frac{B({\mR}(Q))}{Q} \;,
\end{equation} 
where $B({\mR}(Q))$ is a dimensionless function of $\mR$ . This
can be rearranged to obtain, 
\begin{equation} 
  \frac{\rd{\mR}(Q)}{\rd\,{\ln}Q}=B({\mR}(Q))\;.  
\end{equation} 
This is a separable first-order differential equation. In order to solve it
we will need to impose a boundary condition, and to know something of the
behaviour of $B({\mR})$. We shall assume that ${\mR}$ has the
perturbative expansion, 
\begin{equation}
  {\mR}=a\left(1+{\sum_{n>0}}{r_n}{a}^{n}\right)\;, 
\end{equation}
where $a{\equiv}{\as({\mu})}/{\pi}$ is the RG-improved coupling. The
form of expansion in Eq.(5) can always be arranged by suitably scaling the
observable and raising to an appropriate power. The required boundary
condition will be given by asymptotic freedom, that is
${\mR}({\infty})=0$. Integrating Eq.(4) one then obtains
\begin{equation}
  {\ln}\frac{Q}{{\Lambda}_{\mR}}={\int_{0}^{{\mR}(Q)}}\frac{dx}{B(x)}\;+{\kappa}\;.
\end{equation} The constant of integration has been split into
${\ln}{\Lambda}_{\mR}+{\kappa}$, where ${\Lambda}_{\mR}$ is a finite
dimensionful scale specific to the observable ${\mR}$, and ${\kappa}$ is
a universal infinite constant needed to implement ${\mR}({\infty})=0$. To
determine $\kappa$ we need to know the behaviour of $B(x)$ around $x=0$.
Returning to the perturbative series of Eq.(5) we note that the coupling $a$
satisfies the beta-function equation, 
\begin{equation} 
  {\mu}\frac{\partial a}{\partial{\mu}} =
  {\beta}(a)=-b{a}^{2}\left(1+ca+{\sum_{n>1}}{c_n}{a}^{n}\right)\;, 
\end{equation} 
where $b=(33-2{N_f})/6$ , and $c=(153-19{N_f})/12b$, are the first two
coefficients of the beta-function for SU(3) QCD with $N_f$ active (massless)
flavours of quark. They are universal, whereas the subsequent coefficients
$c_i, (i>1)$ are scheme-dependent. In fact, as shown by Stevenson
\cite{r20}, the non-universal beta-function coefficients can be used to label
the renormalisation scheme (RS), together with the renormalisation scale.  If
we set ${\mu}=Q$ in Eq.(5), and differentiate with respect to ${\ln}Q$
term-by-term using the beta-function equation (7), we can obtain
$B({\mR}(a))$ as a power series in $a$, finally if we invert the series
${\mR}(a)$ in Eq.(5) to obtain $a({\mR})$ as a power series in
${\mR}$, we can obtain $B({\mR})$ as a power series in ${\mR}$ .
One finally finds for the series expansion of $B(x)$ around $x=0$, 
\begin{equation} 
  B(x)=-b{x}^{2}\left(1+c{x}+{\sum_{n>1}}{\rho}_{n}{x}^{n}\right)\;.
\end{equation} 
The first two coefficients $b$, $c$, are the universal beta-function
coefficients.  The higher terms ${\rho}_{i}, (i>1)$ are renormalisation
scheme (RS)-invariant, and $Q$-independent, combinations of the ${r_i}$ and
${c_i}$.  The first two are \cite{r14,r18,r20}
\begin{eqnarray}
{\rho}_{2}&=&{c_2}+{r_2}-{r_1}{c}-{r}_{1}^{2}  \nonumber \\
{\rho}_{3}&=&{c_3}+2{r_3}-4{r_1}{r_2}-2{r_1}{\rho}_{2}-{r}_{1}^{2}c+2{r}_{1}^{3}\;.
\end{eqnarray}
Knowledge of ${\rho}_{n}$ requires a complete ${\rm{N}}^{n}$LO
perturbative calculation, that is a calculation of the $r_{i}$ for
$i=1,2,{\ldots},n$ and the ${c_i}$ for $i=2,3,{\ldots},n$ , in some
renormalisation scheme, for instance \msb.  The fact that Eq.(8)
has the same form as the beta-function equation (7) follows from the fact
that there exists an RS in which ${\mR}=a$, i.e. ${r_i}=0, i>0$, and in
this scheme the non-universal beta-function coefficients are
${c_i}={\rho}_{i}$, $i>1$. The existence of this Effective Charge scheme
\cite{r14} is underwritten by the algebraic steps above from which (8) can be
directly derived.  Armed with knowledge of the form of $B(x)$ around $x=0$ we
see that the infinite constant of integration ${\kappa}$ will be of the form
\begin{equation} 
  {\kappa}=-{\int_0^C}\frac{dx}{K(x)}, 
\end{equation}
where $K(x)$ must be such that the singularity of $1/B(x)$ at $x=0$ in (6) is
cancelled. This implies from (8) that 
\begin{equation} 
  K(x)= -b{x}^{2}(1+cx+{\Delta}(x)), 
\end{equation} 
where ${\Delta}(x)$ is only constrained by the requirement that
${\Delta}(x)/{x^2}$ is finite as $x{\rightarrow}0$. Different choices of the
upper limit of integration, $C$, and the function ${\Delta}(x)$, can be
absorbed into the dimensionful constant ${\Lambda}_{\mR}$. Convenient
choices are $C={\infty}$ and ${\Delta}(x)=0$. With these choices (6) can be
re-written as,
\begin{equation} 
  b{\ln}\frac{Q}{{\Lambda}_{\mR}}=
  {\int_{{\mR}(Q)}^{\infty}}\frac{dx}{{x^2}(1+cx)}+
  {\int_0^{{\mR}(Q)}}{dx}\left[\frac{b}{{B}(x)}+\frac{1}{{x^2}(1+cx)}\right]\;.
\end{equation} 
The first integral on the r.h.s. of (12) gives 
\begin{equation} 
  F({\mR}) {\equiv}
  \frac{1}{{\mR}}+c{\ln}\left[\frac{c{\mR}}{1+c{\mR}}\right]\;.
\end{equation} 
Denoting the second integral by $G({\mR})$ we have 
\begin{equation} 
  b{\ln}\frac{Q}{{\Lambda}_{\mR}}=F({\mR})+G({\mR})\;. 
\end{equation}
The desired inverse function ${\Phi}^{-1}$ of (2) can then be obtained by
exponentiating (14), which gives 
\begin{equation}
  {\mathcal{F}}({\mR}(Q)){\mG}({\mR}(Q))=\frac{{\Lambda}_{\mR}}{Q}\;
\end{equation} 
where ${\mathcal{F}}$ is the universal function 
\begin{equation} 
{\mathcal{F}}({\mR}){\equiv}{e}^{-F({\mR})/b}={e}^{-1/b{\mR}}{(1+1/c{\mR})}^{c/b}\;,
\end{equation} 
and 
\begin{equation} 
{\mG}({\mR}) {\equiv}
{e}^{-G({\mR})/b} =
{\exp}\left[-{\int_{0}^{{\mR}(Q)}}{dx}\left(\frac{1}{B(x)} 
    + \frac{1}{b{x}^{2}(1+cx)}\right)\right]\;. 
\end{equation}
If only a NLO perturbative calculation has been completed then our state of
knowledge of $B(x)$ is $B(x)=-b{x}^{2}(1+cx)$ since the NNLO and higher RS
invariants ${\rho}_{2},{\rho}_{3},{\ldots}$ of (8) will be unknown. From (17)
we then have ${\mG}({\mR})=1$. We finally need to relate the
observable-dependent constant of integration ${\Lambda}_{\mR}$ which
arose on integrating (4), to the {\it universal} dimensional transmutation
constant which depends only on the subtraction procedure used to remove the
ultraviolet divergences, \lmsb\ for instance.  Fortunately it turns out that
they can be related {\it exactly} given only a one-loop (NLO) perturbative
calculation of the observable. To see this we begin by noting that on
rearranging (14) and taking the limit as $Q{\rightarrow}{\infty}$ , we obtain
an operational definition of ${\Lambda}_{\mR}$, 
\begin{equation}
  {\Lambda}_{\mR}={\lim_{Q\rightarrow{\infty}}}Q{\exp}(-F({\mR}(Q))/b)\;.
\end{equation}
We have used the fact that $G(0)=0$ together with asymptotic freedom. If
we denote by $a(Q)$ the \msb\ coupling with ${\mu}=Q$ we see that
it will satisfy the beta-function equation (7), of the same form as (4) for
${\mR}$, with ${\beta}_{\msb}(a)$ replacing $B({\mR})$.
This may be integrated following the same steps as above. The constant of
integration ${\Lambda}_{\mR}$ will be replaced by
${\tilde{\Lambda}_{\msb}}$, and the coefficients ${\rho}_{i}$ by the
\msb\ beta-function coefficients ${c}_{i}^{\msb}$. Again
choosing $C={\infty}$ and ${\Delta}(x)=0$, we arrive at 
\begin{equation} 
{\tilde{\Lambda}}_{\msb}={\lim_{Q\rightarrow{\infty}}}Q{\exp}(-F(a(Q))/b)\;.
\end{equation} 
From the perturbative expansion of ${\mR}$ in (5) we will
have 
\begin{equation} 
{\mR}(Q)={a(Q)}+{r}{(a(Q))}^{2}+{\ldots}\;, 
\end{equation} 
where we have defined for
convenience ${r}{\equiv}{r}_{1}^{\msb}({\mu}=Q)$, as the notation
suggests $r$ is Q-independent. It is then straightforward to show that as
$Q{\rightarrow}{\infty}$ 
\begin{equation}
  F({\mR}){\approx}F(a)-{r}+{\ldots}\;, 
\end{equation}  
where the ellipsis denotes terms which vanish as ${Q}{\rightarrow}{\infty}$.
Inserting this result into (18), and comparing with (19), one finally finds
\begin{equation} 
  {\Lambda}_{\mR}={e}^{r/b}{\tilde{\Lambda}}_{\msb}\;,
\end{equation}
for the promised exact relation between the observable-dependent and
universal ${\Lambda}$'s. The tilde over ${\Lambda}$ is to draw attention to
the fact that the above choice of infinite integration constant ${\kappa}$
does not accord with the standard choice \cite{r21}, which is based on an
expansion of $a$ in inverse powers of ${\ln}({\mu}^{2}/{\Lambda}^{2})$.  This
definition corresponds to translating ${\kappa}$ by the finite shift
$c\:\!{\ln}({b}/{2c})$, so that the standard \lmsb\ is related to
${\tilde{\Lambda}}_{\msb}$ by 
\begin{equation}
  {\tilde{\Lambda}}_{\msb}={\left(\frac{2c}{b}\right)}^{-c/b}{\Lambda}_{\msb}\;.
\end{equation} 
Finally assembling all this we arrive at the desired relation between the
universal dimensional transmutation parameter \lmsb\ and
the QCD observable ${\mR}$, 
\begin{equation}
\lmsb =
Q{\mathcal{F}}({\mR}(Q)){\mG}({\mR}(Q)){e}^{-r/b}{(2c/b)}^{c/b}\;. 
\end{equation} 
Notice that all dependence on the subtraction convention chosen to remove
ultraviolet divergences resides in the single factor ${e}^{-r/b}$, the
remainder of the expression being independent of this choice. This is
equivalent to the observation of Celmaster and Gonsalves \cite{r22} that
${\Lambda}$'s with different subtraction conventions can be {\it exactly}
related
given a one-loop (NLO) calculation. \\
As noted above if only a NLO calculation has been performed then the state of our
knowledge of the function $B({\mR})$ in (4) is $B({\mR})=-b{\mR}^{2}(1+c{\mR})$,
and then from (17) ${\mG}({\mR})=1$. So at NLO the {\it best} we can do in
extracting \lmsb\ from the data is
\begin{equation} 
\lmsb = Q{\mathcal{F}}({\mR}(Q)){e}^{-r/b}{(2c/b)}^{c/b}\;.
\end{equation} 
If two-loop (NNLO) and higher-order perturbative calculations are available then
${\mG}({\mR})$ will differ from unity by calculable corrections.
One can expand ${\mG}({\mR})$ as a power series in $\mR$,
\begin{equation} 
{\mG}({\mR})=1-\frac{{\rho}_{2}}{b}{\mR}+O({\mR}^{2})+{\ldots}\;,
\end{equation} 
where ${\rho}_{2}$ is the NNLO RS-invariant defined in (9). Alternatively $G({\mR})$ can
be expanded in the exponent as a power series in ${\mR}$ by expanding the integrand
in (17), to give
\begin{equation} 
{\mG}({\mR})={\exp}[(-{\rho}_{2}{\mR}+O({\mR}^{2})+{\ldots})/b]\;.
\end{equation} 
One could also evaluate the integral in (17) numerically with $B(x)$
truncated, so that at NNLO for instance $B(x)=-b{x^2}(1+cx+{\rho}_{2}{x^2})$.\\

Focussing now on the NLO case where ${\mG}=1$ we note that (25) can be
inverted to give
\begin{eqnarray} 
{\mR}(Q)&=&-\frac{1}{c[1+W(z(Q))]}
\nonumber \\
z(Q)&{\equiv}&-\frac{1}{e}{\left(\frac{Q}{{\Lambda}_{\mR}}\right)}^{-b/c}\;,
\end{eqnarray} 
where $W$ is the Lambert $W$-function \cite{r24,r25} defined implicitly by
the equation $W(z) {\exp}(W(z)) = z$.  To be consistent with asymptotic
freedom it is actually the $W_{-1}$ branch of the function which is required
\cite{r25}.  Eq.(28) is equivalent to the two-loop \msb\ coupling with scale
${\mu}={e}^{-r/b}Q$, and in this scheme ${r_1}=0$. This scheme is sometimes
referred to as the ``Fastest Apparent Convergence'' (FAC) scheme \cite{r20},
and is equivalent to Grunberg's Effective Charge approach at NLO
\cite{r13,r14}.  Crucially, we have derived (28) without having to argue for
a specific choice of scale. Starting from the form of $Q$-dependence of
${\mR}$ implied by dimensional analysis in (3), we simply solved this
differential equation applying asymptotic freedom as a boundary condition. To
define the required infinite constant of integration we needed to know the
series expansion of $B(x)$ around $x=0$, Eq.(8), whose form is completely
scheme-independent, and we thus arrived at Eq.(14). The constant of
integration ${\Lambda}_{\mR}$ could then be exactly related to the universal
dimensional transmutation parameter \lmsb\ associated with use of \msb\ 
subtraction to remove ultraviolet divergences, given a NLO calculation of
$r$, as in Eq.(22). In all of this the renormalised coupling $a$ only ever
appeared in intermediate steps, playing, as neatly expressed in \cite{r15},
``the role of a conjuror's handkerchief- now you see it, now you don't!''.
This, of course, begs the question as to what is special about the Effective
Charge (FAC) scheme, and why other choices of scale ${\mu}$ do not provide
equally valid predictions for ${\mR}$. The key is to identify the way in
which the $Q$-dependence of ${\mR}(Q)$ arises. In the construction above it
is built automatically by integration of (3), but how does it arise from the
 perturbation series in Eq.(5) ?  The crucial observation is that the
perturbative coefficients ${r_i}$ contain ultraviolet logarithms of $Q$. To
see this we can rearrange Eq.(22) to obtain
\begin{equation} 
r=b{\ln}\frac{Q}{{\tilde{\Lambda}}_{\msb}}-b{\ln}\frac{Q}{{\Lambda}_{\mR}}\;,
\end{equation} 
or for a general choice of (\msb) scale $\mu$ ,
\begin{equation} 
{r_1}({\mu}) =
b{\ln}\frac{\mu}{{\tilde{\Lambda}}_{\msb}}-b{\ln}\frac{Q}{{\Lambda}_{\mR}}\;.
\end{equation} 
Thus $r_1$ is a difference of a scheme-dependent logarithm involving $\mu$
and a ``physical'' scheme-independent ultraviolet logarithm involving $Q$. In
RG-improvement as customarily applied one chooses ${\mu}=xQ$, and so the
renormalised coupling $a(xQ)$ is $Q$-dependent. The perturbative coefficients
${r_i}({\mu}=Q)$, however, are $Q$-independent. The consistency of
perturbation theory means that higher coefficients $r_i$ are
${i}^{\rm{th}}$-order polynomials in $r_1$ with $Q$-independent, but
scheme-dependent coefficients \cite{r11,r12,r20}. Rearrangement of Eq.(9),
for instance, gives
\begin{eqnarray}  
{r_2} &=& {r}_{1}^{2}+{r}_{1}c+({\rho}_{2}-{c_2}) 
\nonumber \\
{r_3} &=&
{r}_{1}^{3}+\frac{5}{2}c{r}_{1}^{2}+(3{\rho_2}-2{c_2}){r_1} +
(\frac{{\rho}_{3}}{2}-\frac{c_3}{2})\;.
\end{eqnarray} 
From Eq.(30) we see that ${r_1}({\mu}=xQ)$ is $Q$-independent, and it follows
from (31) that ${r_2},{r_3},{\ldots}$, are therefore $Q$-independent too.
Thus the $Q$-dependence comes {\it entirely} from the renormalised coupling,
and is hence dependent on the unphysical renormalisation scheme parameter
$x$. In contrast the idea of Complete RG-improvement (CORGI) \cite{r11,r12}
is to keep $\mu$ strictly {\it independent} of $Q$. In which case the
$Q$-dependence is built entirely by the ultraviolet logarithms of $Q$
contained in the perturbative coefficients.  Standard NLO fixed-order
perturbation theory is then manifestly inapplicable, since one has
\begin{equation} 
{\mR}_{NLO}=a({\mu})+\left(b{\ln}\frac{{\mu}}{{\tilde{\Lambda}}_{\msb}}-
b{\ln}\frac{Q}{{\Lambda}_{\mR}}\right){(a({\mu}))}^{2}\;.
\end{equation} 
With $\mu$ constant, asymptotic freedom only arises if {\it all} the
RG-predictable UV logarithms are resummed to {\it all-orders}. Given only a
NLO calculation the RS-invariants ${\rho}_{i}, (i>1)$ are unknown, and so the
resummation of the RG-predictable UV logarithms corresponds to setting the
${\rho}_{i}$ to zero in Eq.(31). If we further set $c=0$ and ${c_i}=0, (i>1)$
, to simplify the analysis then the all-orders sum of RG-predictable terms
reduces to a geometric series,
\begin{equation} 
{\mR}=a({\mu})\left(1+{\sum_{n>0}}{(a({\mu}){r}_{1}({\mu}))}^{n}\right)\;.
\end{equation} 
With these simplifications 
 we will have $a({\mu})=1/b{\ln}({\mu}/{\tilde{\Lambda}}_{\msb})$,
and using Eq.(30) and summing the geometric series one obtains,
\begin{equation} 
{\mR}(Q)=a({\mu})/\left[1-\left(b{\ln}\frac{{\mu}}{{\tilde{\Lambda}}_{\msb}}
-b{\ln}\frac{Q}{{\Lambda}_{\mR}}\right)a({\mu})\right]=
1/b{\ln}(Q/{\Lambda}_{\mR})\;,
\end{equation} 
in which the unphysical $\mu$-dependence has cancelled between $a({\mu})$ and the
$\mu$-dependent logarithms contained in ${r}_{1}({\mu})$. In the realistic case
with nonzero $c$ and $c_i$ the simple logarithm of $Q/{\Lambda}_{\mR}$ is replaced by
the Lambert $W$-function of Eq.(28). The key point is that the all-orders
CORGI improvement can be carried out with {\it any} choice of $\mu$ to yield a
$\mu$-independent result. One has therefore directly traded 
unphysical $\mu$-dependence for the physical $Q$-dependence. \\

To further emphasise the connection of the suggested direct
extraction of \lmsb\ with the standard approach we can consider the following result for ${\Lambda}_{\overline{MS}}({r_1},{\mR})$, which we
define to be the value of \lmsb\ obtained by fitting
a NLO perturbative calculation in a scheme corresponding to the
NLO coefficient $r_1$, to the data ${\mR}$
. Notice that $r_1$ completely labels the scheme
at NLO. We can directly convert $r_1$ into the \msb\ scale
$\mu$ since from (29) and (30) we have 
\begin{equation} 
{r_1}=r+b{\ln}\frac{{\mu}}{Q}\;.
\end{equation} 
It is then straightforward to derive the result \cite{r18}
\begin{equation} 
{\lmsb ({r_1},{\mR}) = \frac{{\exp}[f({r_1},{\mR})/b]}
{{\mG}({\mR})}\;\lmsb}\;,
\end{equation} 
where $f({r_1},{\mR})$ is given by
\begin{equation} 
f({r_1},{\mR}){\equiv}F({\mR})-F\left(\frac
{-1+\sqrt{1+4{r_1}{\mR}}}{2{r_1}}\right)+{r_1}\;.
\end{equation} 
In the CORGI approach ${r_1}=0$ and we have $f(0,{\mR})=0$, so that
the value of \lmsb\ obtained is $\lmsb/{\mG}({\mR})$, as
expected comparing (24) and (25). Thus to the extent that
${\mG}({\mR}){\approx}1$ we obtain the {\it actual} value of
\lmsb. As we have argued the estimate
${\mG}({\mR}){\approx}1$ is the {\it best} we can do given
only a NLO calculation since we are in complete ignorance of the deviations
of ${\mG}$ from unity, which will depend on the NNLO RS-invariant
${\rho}_{2}$ of Eq.(9). Another way of saying this is that at asymptotic
values of $Q$ Eq.(25) will hold, and that the deviation of
${\mG}({\mR})$ from unity provides an operational definition
of how far from asymptotia we are, at $Q={M_Z}$, say. The scatter of the
\lmsb\ values for different observables obtained from Eq.(25) thus provides
unambiguous information about the size of sub-asymptotic effects
(uncalculated NNLO and higher perturbative corrections and power
corrections). Variation of the renormalisation scale taking ${\mu}=xQ$ with
the ``physical scale'' $x=1$ giving a central value merely serves to confuse
matters. For instance taking $r=10$ and ${\mR}=0.05$, values typical
of jet observables at $Q={M_Z}$, we find ${\exp}[f(10,0.05)/b]=2.44$, and so
using the ``physical scale'' the value of \lmsb\ extracted will be $2.44
\lmsb/{\mG}({\mR})$. This will accurately determine \lmsb\ if
it fortuitously happens that ${\mG}({\mR}){\approx}2.44$ . We
have, of course, no reason to suppose that ${\mG}$ differs from unity
to such a drastic extent, or correspondingly that the effect of uncalculated
NNLO and higher-order perturbative corrections, and possible power
corrections should be so large. Varying the scale simply introduces an extra
{\it known} factor into the determination of \lmsb, which, {\it other things
  being equal}, i.e. if ${\mG}{\approx}1$, will give values very
different from
the true one.\\
In all this discussion we have considered strictly massless quarks.  In
reality the dimensional transmutation parameter has a dependence on the
number of active quark flavours, $N_f$, so really we have
$\Lambda_{\msb}^{({N_f})}$. The NLO correction $r$ and the
universal beta-function coefficients $b$, $c$, in Eqs.(24,25) also depend on
$N_f$. Transformation between $\Lambda_{\msb}^{({N_f})}$ for
different values of $N_f$ can be effected using the standard apparatus of the
decoupling theorem augmented with a matching condition \cite{r25a}. The
matching condition has now been computed to the three-loop level \cite{r25b}.
For all of our fits in Section 4, ${N_f}=5$ will be the active number of
flavours, and we shall be extracting $\Lambda_{\msb}^{(5)}$.
\section*{3 Definition of the Jet Observables}

We restrict ourselves throughout to infrared safe observables and make all
definitions in the centre-of-mass frame, with all sums running over $N$ final
state particles. We begin by reviewing the various possibilities
for
clustering particles to form jets. Given a particular jet measure, $y_{ij}$
the following algorithm is common to all,
\begin{enumerate}
\item Define a resolution parameter, $y_{cut}$.
\item For every pair of hadrons, $h_i$ and $h_j$, evaluate the jet measure,
  $y_{ij}$. 
\item If the smallest occurrence of this quantity is less than the resolution
  parameter (i.e. $\mbox{min}(y_{ij}) < y_{cut}$) combine the corresponding
  hadron momenta, $p_i$ and $p_j$ into that of a pseudo-particle, $p_{ij}$
  according to a recombination prescription.
\item Repeat steps 2--4 until all hadrons and pseudo-hadrons have jet
  measures greater than the resolution parameter. What remains are then
  denoted \emph{jets}.
\end{enumerate}

By introducing a jet resolution parameter, $y_{cut}$, we have made our
definition of a jet intrinsically infrared safe. Increasing its value permits
a greater number of clusterings and thus few jet events are identified.
Likewise, decreasing its value finally results in all final state hadrons
being assigned to separate jets. Within the theoretical framework, such small
values probe deeply into the infrared region and thus require a thorough
treatment of hadronisation. Jet measures are typically normalised by the
total visible energy of the hadronic event, $E_{\mathrm{vis}}$, to give a
dimensionless quantity.
For a description of the multitude of different algorithms with  a discussion of their merits
see \cite{Moretti:1998qx}.  For this analysis we shall restrict ourselves to
the JADE, Durham and Geneva jet finding measures applied
by the various experimental collaborations.\\

The first jet measure to be proposed was by the JADE collaboration
\cite{Bethke:1988zc} and simply uses,
\begin{equation}
  \label{eq:JADE}
  y_{ij}^J = \frac{2E_i E_j(1-\cos\theta_{ij})}{E_{\mathrm{vis}}^2}
  \approx  \frac{M_{ij}^2}{E_{\mathrm{vis}}^2} ,
\end{equation}
where $E_l$ denotes the energy of a hadron, $l$, in the centre-of-mass frame
and $\theta_{lm}$ is the opening angle of the pair under consideration. In
the massless limit this measure corresponds to their invariant mass,
$M_{ij}^2$.  Having defined the jet measure we are still at liberty to define
the procedure for recombining two hadrons into a pseudo-hadron. There are
four immediately obvious possibilities, denoted the $E$, $E0$, $P$ and $P0$
schemes. In all cases the subscript $k$ denotes the pseudo-particle created by
particles $i$ and $j$.
In the E scheme, we combine two particles according to their four-momenta,
\begin{equation}
  \label{eq:E-scheme}
  p_k = p_i + p_j.
\end{equation} 
Energy and momentum are explicitly conserved in this scheme.
In the so-called E0 Scheme 
 the three-momenta of the pseudo-particle is rescaled to give
it zero invariant mass,
\begin{eqnarray}
  E_k &=& E_i + E_j,\\
  {\bf p}_k &=& \frac{E_k}{|{\bf p}_i+{\bf p}_j|} ({\bf p}_i+{\bf p}_j).
\end{eqnarray}
As a result the total momentum sum of the event is not conserved.
Conversely in the P Scheme we may conserve the total momentum of the event at the expense of
the total energy conservation using
\begin{eqnarray}
  {\bf p}_k &=& {\bf p}_i + {\bf p}_j,\\
  E_k &=& |\bf{p}_k|.
\end{eqnarray}
Lastly we introduce a variation of the $P$ scheme, the P0 Scheme, by altering the jet measure
such that after recombination, the total visible energy is changed such that,
\begin{equation}
  E_{\mathrm{vis}} = \sum_k E_k.
\end{equation}

Unfortunately the JADE jet measure turns out to introduce spurious
clusterings in certain circumstances whereby a resultant jet is formed in a
direction lacking any approximately collinear initial hadrons. This
translates into theoretical problems when attempting to perform large
infrared logarithm resummations where these correlations spoil the property
of exponentiation in the two-jet limit \cite{Brown:1992hx}.
A subsequent attempt to suppress artificial recombinations within the jet
clustering and hence improve its theoretical properties was suggested by
Dokshitzer et al. \cite{Catani:1991hj}, termed the Durham or
${k_\perp}$-algorithm. It uses the minimum relative transverse momenta
of two hadrons in the small angle limit,
\begin{equation}
  \label{eq:Durham-alg}
  y^D_{ij}=\frac{2\,\min(E_i^2,E_j^2)(1-\cos\theta_{ij})}{E_{\mathrm{vis}}^2}.
\end{equation}
This form of clustering reduces the number of spurious recombinations and
permits a straightforward theoretical implementation. As such
it has now become the standard algorithm in use. We use the E scheme
recombination. 

Lastly we consider a variant termed the Geneva algorithm proposed by
Bethke et al.  \cite{Bethke:1992wk} that also attempts to
reduce the spurious mis-clusterings of the Jade algorithm using the measure,
\begin{equation}
  \label{eq:Geneva-alg}
  y^G_{ij}=\frac{8}{9}\frac{E_iE_j(1-\cos\theta_{ij})}{(E_i+E_j)^2}.  
\end{equation}
In contrast to the previous two proposals, the Geneva algorithm does not
depend on the energy of the event, and has a preference to combine soft
particles with hard ones. This in turn reduces the correlations between soft
gluons when performing infrared logarithm resummations. We also use the E
scheme for recombination. 

With the jet finding algorithms in place we may now determine the $n$-jet
rates ($R_n(y_{cut})$) by the fraction of events with $n$ resultant jets after
clustering. We may then define the jet transition parameters, $y_{n\;n+1}$
that corresponds to the value of $y_{cut}$ where an event changes from
$(n+1)$-jet-like to $n$-jet-like.\\ 

We now turn to Event Shape Variables.
Many of these variables are related and can be broadly categorised as
follows. Note that they will in general contain both three and four-jet-like
quantities. 
Thrust ($T$) is defined by maximising the net longitudinal momentum
of final state particles along the direction of a thrust axis
\cite{Brandt:1964sa,Farhi:1977sg},
\begin{equation}
  \label{eq:thrustdef}
T=\mbox{max}\frac{{\sum_{i}}|{\bf{p}}_{i}\cdot{\bf{n}}_{T}|}
{{\sum_{i}}|{\bf{p}}_{i}|}\;,
\end{equation}
where ${\bf{p}}_{i}$ denotes the final state particle momenta, and
${\bf{n}}_{T}$ denotes the unit vector in the direction of the thrust axis,
to be determined by maximising the above quotient.  Defining for convenience
${\tau}{\equiv}1-T$ , we find that $\tau$ varies between zero, for two
back-to-back final state partons, up to a maximum of ${\tau}=\frac{1}{2}$ for
spherical (isotropic) events. For planar events with three final-state
partons, one finds a maximum value of ${\tau}=\frac{1}{3}$ corresponding to a
``Mercedes Benz'' configuration. Two further variants, thrust-major
(${T}_{\mathrm{maj}}$) and thrust-minor (${T}_{\mathrm{min}}$) can
be defined. In ${T}_{\mathrm{maj}}$ the thrust axis ${\bf{n}}_{T}$ is replaced in
Eq.(\ref{eq:thrustdef}) by ${\bf{n}}_\mathrm{maj}$, which maximises the sum of
momenta transverse to the thrust axis.  In ${T}_\mathrm{min}$ it is replaced by an
axis ${\bf{n}}_\mathrm{min}$ which is the vector cross product of ${\bf{n}}_{T}$ and
${\bf{n}}_\mathrm{maj}$. One can then define the oblateness $O$ by
\cite{Barber:1979yr}
\begin{equation}
O{\equiv}{T}_\mathrm{maj}-{T}_\mathrm{min}\;.
\end{equation}

Events can also be divided into two hemispheres (${\bf H}_a,{\bf H}_b$) by a
plane perpendicular to ${\bf{n}}_{T}$. We may then calculate the normalised
invariant mass of each hemisphere ($x = a, b$) \cite{Clavelli:1979md},
\begin{equation}
  \label{eq:mass-def}
  \rho_x = \frac{M_x^2}{E_{\mathrm{vis}}^2} = \frac{1}{E_{\mathrm{vis}}^2}\left( \sum_{{\bf p}_k
  \in {\bf H}_x} p_k \right)^2.
\end{equation}
This permits the possibility of four obvious combinations giving rise to 
\begin{eqnarray}
  \label{eq:en-mass}
  \rho_{T} &=& \rho_a + \rho_b,\\
  \rho_{D} &=& |\rho_a - \rho_b|,\\
  \rho_{H} &=& \mbox{max}(\rho_a,\rho_b),\\
  \rho_{L} &=& \mbox{min}(\rho_a,\rho_b),
\end{eqnarray}
which correspond to the sum of jet masses, the difference of
  jet masses, the heavy jet mass , and the light jet mass,
respectively.  To lowest order in perturbative QCD, and assuming massless
quarks, thrust and heavy jet mass are related by ${\tau}={\rho_H}$
\cite{Kunszt:1989km}.

Other variants on thrust and jet masses are the jet broadening
measures proposed in \cite{Catani:1992jc}.  In each of the above hemispheres
$a$ and $b$ one forms a jet broadening, $B$, by summing over the particles in that
hemisphere,
\begin{equation}
B_x =\frac{ \sum_{{\bf p}_k \in {\bf H}_x} |{\bf{p}}_{i}\times{\bf{n}}_{T}|}
{2{\sum_{{\bf p}_k \in {\bf H}_x}}
|{\bf{p}}_{i}|}\;.
\end{equation}

Once again we may compose a range of variables by the combinations, 
\begin{eqnarray}
  \label{eq:en-broad}
  B_{T} &=& B_a + B_b,\\
  B_{D} &=& |B_a - B_b|,\\
  B_{W} &=& \mbox{max}(B_a,B_b),\\
  B_{N} &=& \mbox{min}(B_a,B_b),
\end{eqnarray}
to make the sum of hemisphere broadening, the difference of
  hemisphere broadenings, the wide hemisphere broadening and the
narrow hemisphere broadening respectively.  For two-parton final
states ${B}_{T}={B}_{W}=B_D=0$, and to lowest order in perturbation theory
${B_T}={B}_{W}=B_D=\frac{1}{2}O=\frac{1}{2}T_{\mathrm{max}}$.

We can also define the so-called $C$ and
$D$-parameters from the eigenvalues of the infra-red safe linear
momentum tensor \cite{Parisi:1978eg,Donoghue:1979vi},
\begin{equation}
{\theta}_{{m}{n}}=\frac{{\sum_i}{p}_{i}^{m}{p}_{i}^{n}/
|{\bf{p}}_{i}|}{{\sum_i}|{\bf{p}}_{i}|}\;,
\end{equation}
where ${p}_{i}^{m}$ is the ${m}$-th component of the three-momentum
${\bf{p}}_{i}$, with $i$ summed over all final state particles. As
defined the tensor has unit trace. The $C$-parameter is then defined
in terms of the eigenvalues of the tensor ${\theta}_{{\mu}{\nu}}$,
${\lambda}_{1},{\lambda}_{2},{\lambda}_{3}$ by,
\begin{equation}
  \label{eq:C-def}
  C=3({\lambda}_{1}{\lambda}_{2}+{\lambda}_{2}{\lambda}_{3}+
  {\lambda}_{3}{\lambda}_{1})\;.
\end{equation}
$C=0$ for back-to-back two parton final states and $C=1$ for spherical
(isotropic) events.  For planar three-parton final states $0 {\leq} C {\leq}
\frac{3}{4}$ where one of the eigenvalues is zero. For values greater than
$\frac{3}{4}$ requires at least four final state particles.

The $D$-parameter is defined by the combination
\begin{equation}
  \label{eq:D-def}
  D=27 \lambda_1\lambda_2\lambda_3,
\end{equation}
and only becomes non-zero for non-planar events (i.e. four or greater final
states particles).\\

Rather than describe an event by a single variable, we may consider inclusive
two-particle correlations. The energy-energy correlation
(\textit{EEC}) \cite{Basham:1978bw,Basham:1979zq,Basham:1978iq} is the
normalised energy-weighted cross section defined in terms of the angle,
$\chi_{ij}$, between two particles $i$ and $j$ in an event,
\begin{equation}
  \label{eq:EEC-def}
  \mathit{EEC}(\chi) \equiv
 \frac{1}{N \Delta\chi} \sum_{\mathrm{events}}
  \int^{\chi+\frac{\Delta\chi}{2}}_{\chi-\frac{\Delta\chi}{2}} \sum_{i,j}
  \frac{E_i E_j}{E_{\mathrm{vis}}^2} \delta(\chi' -\chi_{ij}) \rd \chi',
\end{equation}
where the argument, $\chi$, is the opening angle to be studied for the
correlations, $\Delta\chi$, is the angular bin width and $N$ is the number of
events. The angle $\chi$ can be varied in the range $0^{\circ} \le \chi \le
180^{\circ}$ where the central region ($\chi \sim 90^\circ$) is governed by
hard gluon emission and the extremities ($\chi \sim 0^{\circ}$ and
$180^\circ$), corresponding to collinear and back-to-back configurations, are
expected to be sensitive to hadronisation.  We may further define the
asymmetric energy-energy correlation (\textit{AEEC}) to be
\begin{equation}
  \label{eq:AEEC-def}
  \mathit{AEEC}(\chi) = \mathit{EEC}(180^\circ-\chi) - \mathit{EEC}(\chi),
\end{equation}
where now $\chi$ is within the range $0^\circ$--$90^\circ$.

A recent addition to the set of \epem\, jet observables is the jet
  cone energy fraction
  (\textit{JCEF}) \cite{Ohnishi:1994vp}. Here the
energy within a conical shell of opening angle $\chi$ about the thrust axis
is integrated,
\begin{equation}
  \label{eq:JCEF-def}
  \mathit{JCEF}(\chi) = \frac{1}{N \Delta\chi} \sum_{\mathrm{events}}
  \int^{\chi+\frac{\Delta\chi}{2}}_{\chi-\frac{\Delta\chi}{2}} \sum_{i}
  \frac{E_i}{E_{\mathrm{vis}}} \delta(\chi' -\chi_{i}) \rd \chi',
\end{equation}
where 
\begin{equation}
  \chi_i = \arccos\left(\frac{{\bf p}_i\cdot{\bf n}_T}{|{\bf p}|}\right)
\end{equation}
is the opening angle between a particle, $i$, and the thrust axis vector,
$n_T$ defined to point from the heavy jet mass hemisphere to the light jet
mass hemisphere. The angle, $\chi$, is within the range $0^{\circ} \le
\chi \le 180^{\circ}$ and thus the hard gluon emissions will feature when
$\chi \ge 90^\circ$.

All the observables previously defined are either three or four-jet like.
Additionally, we shall also investigate the two-jet rate.  In the subsequent
analysis we utilise experimental data at the $Z^0$-peak from the SLD
\cite{Abe:1995mf}, ALEPH \cite{Barate:1996fi}, DELPHI
\cite{Abreu:2000ck,Abreu:1996na}, L3 \cite{Adeva:1992gv} and OPAL
\cite{Acton:1992fa} collaborations. 

\section*{4 Direct Extraction of $\boldsymbol{\lmsb}$} 
In order to perform the direct extraction of $\lmsb$
from the data for $\cal{R}$ , using Eq.(25), we shall need to
recast the perturbative expansions for the two, three and four-jet
observables to be considered, so that they have the dimensionless
form assumed in Eq.(5).
For the three-jet-like observables, $\tilde{\mR}_3$, we will in general have
a perturbative expansion at NLO of the form
\begin{equation}
  \tilde{\mR}_3(Q) = A_3 \frac{\as(\mu)}{2\pi} + B_3(\mu)
  \left(\frac{\as(\mu)}{2\pi}\right)^2. 
\end{equation}
The two-jet case is
  equivalent to the three-jet case with the substitution $\tilde{\mR}_3\to
  (1-\tilde{\mR}_2)$.
  Correspondingly four-jet-like observables, $\tilde{\mR}_4$ will have the expansion
\begin{equation}
  \tilde{\mR}_4(Q) = A_4 \left(\frac{\as(\mu)}{2\pi}\right)^2 + B_4(\mu)
  \left(\frac{\as(\mu)}{2\pi}\right)^3,
\end{equation}
where $A_n$ denotes the $\mu$-independent tree-level coefficient of an
$n$-jet-like quantity and $B_n$ the NLO coefficient.  We may then by simple
algebraic manipulation rewrite these in terms of the required dimensionless
quantity as
\begin{equation}
  \label{eq:mr3}
  \mR_3(Q) = \frac{2 \tilde{\mR}_3(Q)} {A_3} =
  \frac{\as(\mu)}{\pi}+ \frac{B_3(\mu)}{2 A_3}
  \left(\frac{\as(\mu)}{\pi}\right)^2
\end{equation}
and 
\begin{equation}
  \label{eq:mr4}
  \mR_4(Q) = \sqrt{\frac{4 \tilde{\mR}_4(Q)} {A_4}} =
  \frac{\as(\mu)}{\pi} + \frac{B_4(\mu)}{4 A_4}
  \left(\frac{\as(\mu)}{\pi}\right)^2. 
\end{equation}

We are now in a position to calculate \lmsb\, from
Eq.(25) by substituting the experimental values of
$\mR(Q)$ and the fundamental quantity, $r_1(\mu=Q)$ which can be read from
Eqs.(\ref{eq:mr3}) and ~(\ref{eq:mr4}). We may then apply this to every
experimental bin, enabling a direct extraction of \lmsb\, across the
kinematic range of the variable. In all cases we use the Monte Carlo programs
{\tt EERAD} \cite{Giele:1992vf} and {\tt EERAD2} \cite{Campbell:1998nn} to
calculate the NLO perturbative coefficients for three and four jet quantities
respectively.

Before we attempt to extract a value for \lmsb\, there are a number of
important issues worth considering. Firstly we must remember that
even though we have defined a set of observables that attempt to reflect the
underlying behaviour of the QCD partons, the effects of hadronisation will
always be present. In some observables this will be more pronounced in
certain regions of phase space resulting in the perturbative prediction
failing to provide a reliable description. A number of Monte Carlo programs
exist \cite{Sjostrand:1994yb,Marchesini:1992ch} that attempt to model this
behaviour and have proved very successful. In addition non-perturbative power
corrections have been studied phenomenologically and have displayed very
positive results too.

\begin{figure}[t]
  \begin{center}
    \epsfig{file=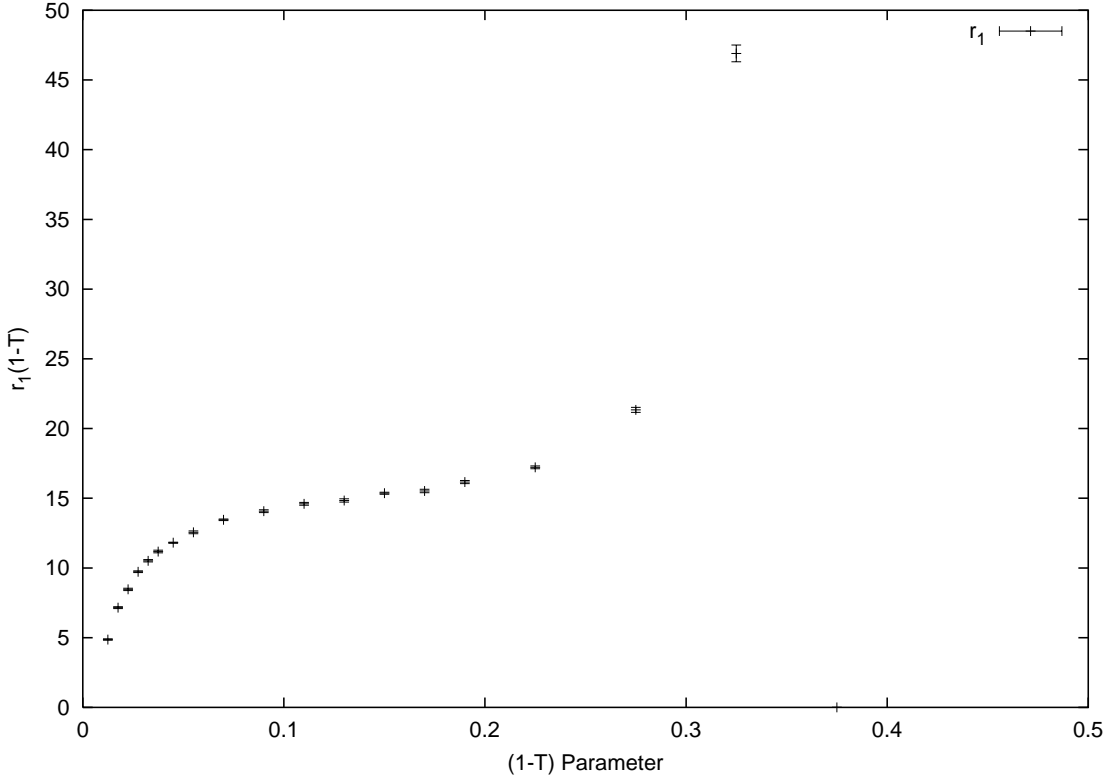,angle=270,width=15cm}
    \caption{The $r_1$ parameter as a function of 1-thrust. It is calculated
      according to the ALEPH  experimental bin sizes with
      errors attributed to the Monte Carlo integration of the LO and NLO
      coefficients only.}
    \label{fig:r1thrust}
  \end{center}
\end{figure}

\begin{figure}[h]
  \begin{center}
    \epsfig{file=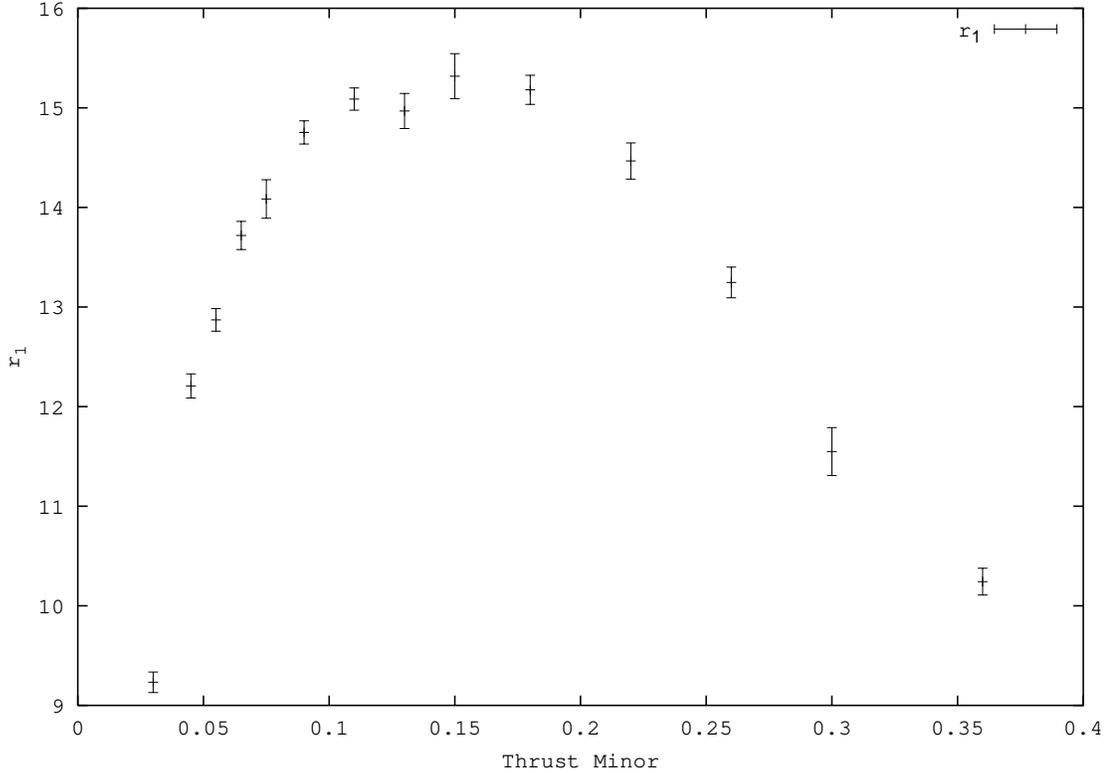,angle=270,width=15cm}
    \caption{The $r_1$ parameter as a function of thrust minor. It is calculated
      according to the DELPHI experimental bin sizes with
      errors attributed to Monte Carlo integration of the LO and NLO
      coefficients only.}
    \label{fig:r1minor}
  \end{center}
\end{figure}

Secondly, from a purely perturbative QCD perspective, semi-inclusive
quantities suffer from large kinematic logarithms at the exclusive boundaries
of phase space. This manifests itself in the $r_1$ parameter, growing in
magnitude typically like $-\ln^2\lambda$ for a variable $\lambda$ that goes
to zero in the two-jet configuration. This drags the value for \lmsb\, to
zero regardless of the true value, and indicates a breakdown in the NLO
approximation since higher-order terms will be enhanced by powers of
logarithms requiring an all-orders resummation. Furthermore, at the opposite
end of the kinematic range we typically encounter a similar problem due to an
end point in phase space.  These occur when a variable goes from being
$n$-jet-like to $(n+1)$-jet-like.  Examples of this are the 1-thrust at
$\frac{1}{3}$ and the C-parameter at $\frac{3}{4}$. Above these values, the
three-jet configurations do not contribute, resulting in the tree level term
vanishing. Clearly this now upsets our definition of $r_1$ since it diverges
in a direction governed by the relative sign difference between LO and NLO in
this limit. These characteristics can be seen in Fig.~\ref{fig:r1thrust}
for the 1-thrust variable and Fig.~\ref{fig:r1minor} for the thrust minor
variable.  A more sophisticated way of handling the end point problem would
be to define a new $r_1$ value according to the ratio of the order $\as^3$
coefficient to order $\as^2$ in the region corresponding to non-zero four-jet
configuration contributions and then smoothly interpolate a value across the
threshold.

These two difficulties must be taken into consideration when attempting to
extract a value for \lmsb\,. For a number of the three-jet quantities,
hadronisation corrected data is analysed, providing a means of reducing that
uncertainty. In these cases the Jetset~7.4 hadronisation model
\cite{Sjostrand:1994yb} was implemented (unless otherwise stated) using
bin-by-bin correction factors with errors estimated via statistical
uncertainty. These factors were
calculated as specified in \cite{Abe:1995mf} and \cite{Abreu:1996na}\\

Having identified the possible difficulties arising we must specify a set of
criteria to perform a direct extraction of \lmsb\,. This should concentrate
on a plateau region of \lmsb\, in the central region of the kinematic range of
the variable. We have adopted the following procedure for specifying the fit
range, 
\begin{enumerate}
\item Decrease the fit range from its maximum value such that all $r_1$
  values lie within a variation of 20\% from the flattest region.
\item Decrease the range further (if necessary) to the region where
  hadronisation corrections are less than 40\%.
\item If more than three points are present, perform a single parameter
  $\chi^2$ fit to a flat line to calculate a value for \lmsb\, with error.
\item Rescale the error according to $\sqrt{\chi^2/N_{df}}$ to obtain a \lmsb\ 
  \,prediction for the given variable and collaboration.
\end{enumerate}

If at any stage there are fewer than four consecutive bins surviving, the jet
variable is considered unsuitable for the analysis. There is, of course, no
guarantee that a variable will have a flat plateau over which to perform the
fit. It may be such that the kinematic boundary effects dominate over the
complete range. In these cases we are forced to disregard the variable.

We have chosen to use an $r_1$ criterion to avoid the problem of large
kinematic logarithms spoiling the fixed order perturbation theory. The
parameter, $r_1$, clearly indicates the region where these logarithms are
dominating the series and hence the breakdown of the NLO approximation. The
value of $r_1$ does not indicate where hadronisation effects may be
considerable though.  Therefore in order to give a proper treatment of the
variables, we should use experimental data that has been corrected for
hadronisation effects. We attempt to include a reasonably flat region across
$r_1$ by allowing a 20\% deviation from flatness (with errors taken into
account). Since the $r_1$ parameter varies smoothly across the kinematic
range, this criterion permits a good measure of flatness. The value of 20\%
is chosen to tolerate minor deviations in $r_1$ in the vicinity of the end
points and any statistical fluctuations from evaluation of the NLO
coefficients which are typically small. The resulting fit range should be
relatively insensitive to small variations in the permitted percentage
deviation. 

If hadronisation corrected data is available, we have adopted the procedure
presented in \cite{Abe:1995mf} for excluding any bins that suffer from
greater than a 40\% correction.

Finally, we adopt a minimum $\chi^2$ test for fitting a flat line to the data
points. The initial error (induced by $\Delta\chi^2=1$ from minimum)
associated with the fit is then scaled by $\sqrt{\chi^2/N_{df}}$ for $N_{df}$
degrees of freedom as promoted in the review of particle physics
\cite{Groom:1998in}. This provides a value of \lmsb\, for each observable
measured by each collaboration. We must then consider how to combine the
values. 

In considering the forthcoming fits, we must be careful not to underestimate
the errors. Dealing with different experiments' measurements of the same
observable will obviously have strong correlations.  Typically the greatest
difference between data sets will be due to statistical errors especially in
the cases without any hadronisation corrections being applied.  A procedure
has been put forward by Schmelling {\cite{Schmelling:1994pz}}, termed the
\textit{method of correlated averages}, to combine correlated data when the
exact correlation matrix is unknown. In this case, it is suggested that the
degree of correlation is set by the $\chi^2/N_{df}$ value of the data set. In
this way we are able to combine any number of correlated data without an
unnatural reduction in the error.  Similarly when combining errors with a
$\chi^2/N_{df}$ greater than one we adopt the standard technique of rescaling
the error by $\sqrt{\chi^2/N_{df}}$ to improve the error estimation according
to the quality of the fit.

The experimental data for the three-jet observables (uncorrected for
hadronisation effects) is taken from \cite{Abe:1995mf} for SLD data,
\cite{Barate:1996fi} for ALEPH data, \cite{Abreu:2000ck} for DELPHI data,
\cite{Adeva:1992gv} for L3 data and \cite{Acton:1992fa} for OPAL data.
Additionally, hadronisation corrected data is applied where available.  In
Figs.~3-15 we give the ${\Lambda}_{\msb}$ plots for a set of thirteen
observables for which hadronisation corrected data is available, and for
which the fits to constant ${\Lambda}_{\msb}$ satisfy the criteria outlined
above. The fit ranges are indicated by vertical dashed lines, and the best
overall fit to ${\Lambda}_{\msb}$ for the data sets considered, by a
horizontal dash-dotted line. These values of ${\Lambda}_{\msb}$ are assembled
in Fig.~17.
\begin{figure}[!htbp]
  \begin{center}
    \epsfig{file=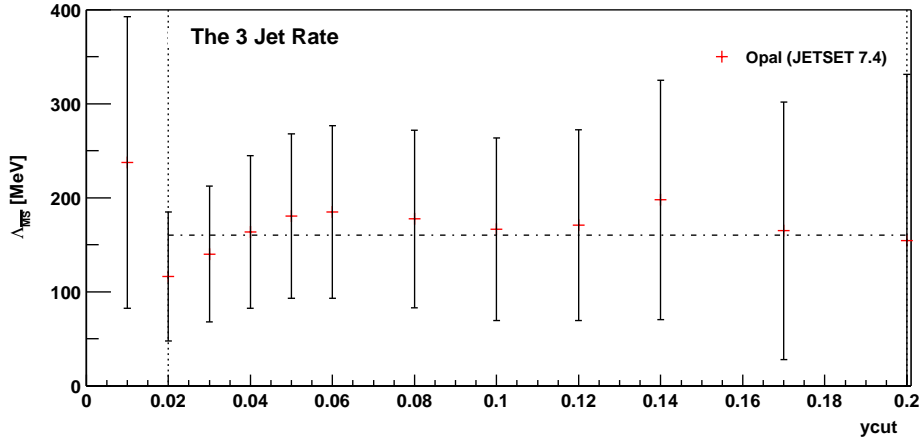,width=\textwidth}    
    \caption[The 3 Jet Rate using OPAL data]{The 3 Jet Rate using OPAL data}
    \label{fig:3jetrate.opl}
  \end{center}
\end{figure}
\begin{figure}[!htbp]
  \begin{center}
    \epsfig{file=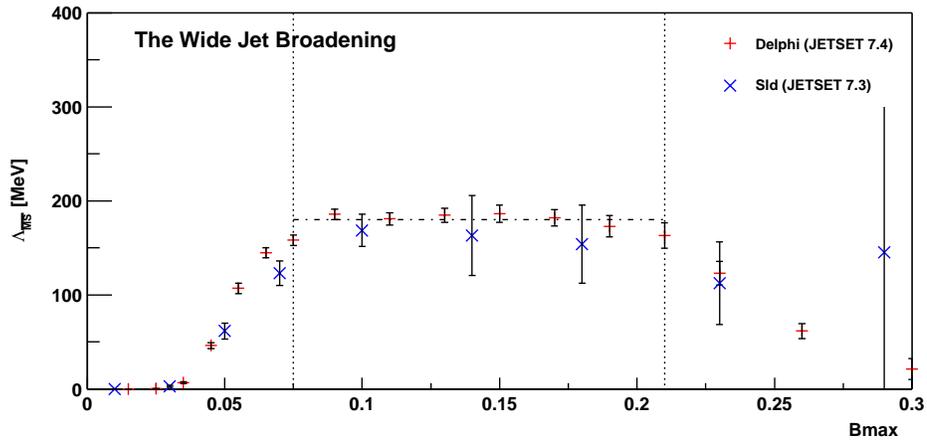,width=\textwidth}    
    \caption{The Wide Jet Broadening using DELPHI and SLD data}
    \label{fig:bmax.dlp}
  \end{center}
\end{figure}
\begin{figure}[!htbp]
  \begin{center}
    \epsfig{file=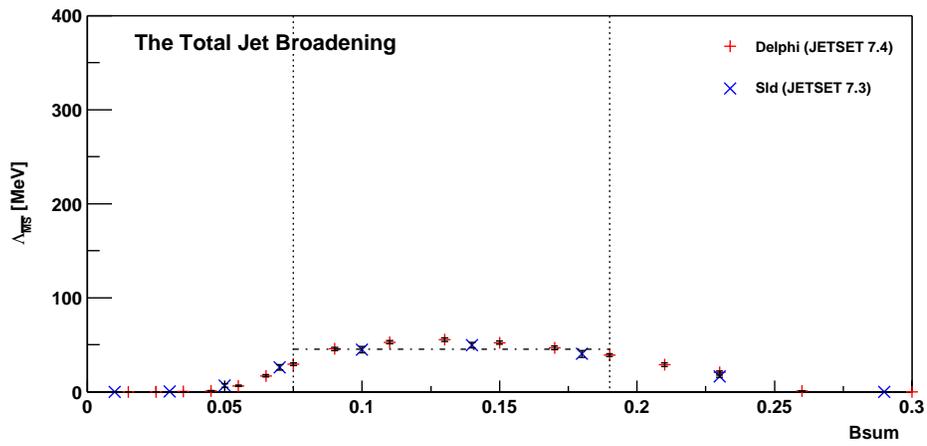,width=\textwidth}    
    \caption{The Total Jet Broadening using DELPHI and SLD data}
    \label{fig:bsum.dlp}
  \end{center}
\end{figure}
\begin{figure}[!htbp]
  \begin{center}
    \epsfig{file=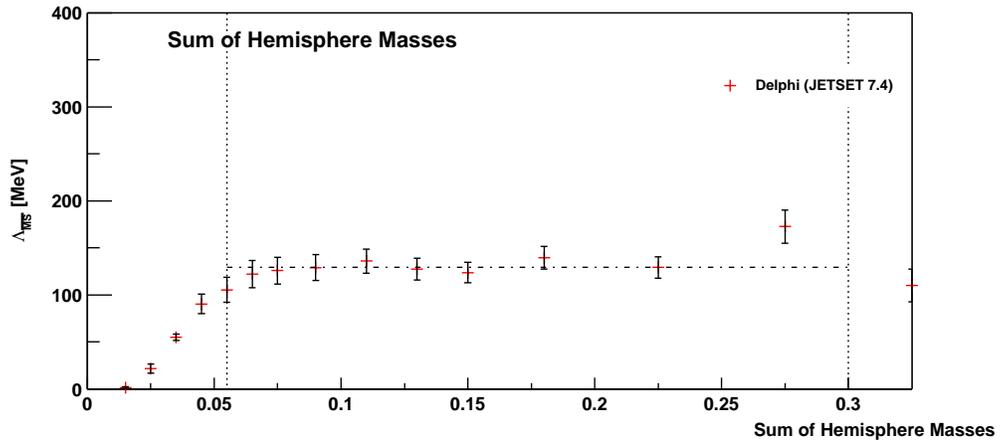,width=\textwidth}    
    \caption{The Sum of Hemisphere Masses using DELPHI data}
    \label{fig:sumjetmass.dlp}
  \end{center}
\end{figure}
\begin{figure}[!htbp]
  \begin{center}
    \epsfig{file=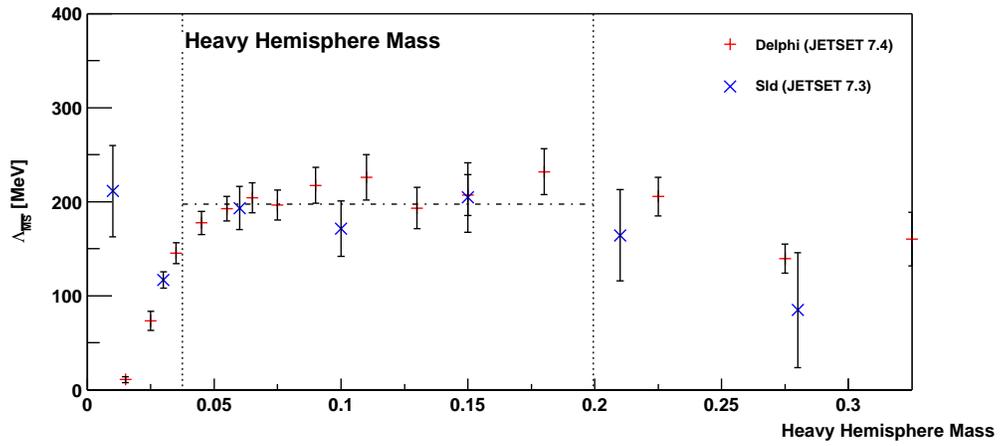,width=\textwidth}    
    \caption{The Heavy Hemisphere Mass using DELPHI and SLD data}
    \label{fig:heavyjetmass.dlp}
  \end{center}
\end{figure}
\begin{figure}[!htbp]
  \begin{center}
    \epsfig{file=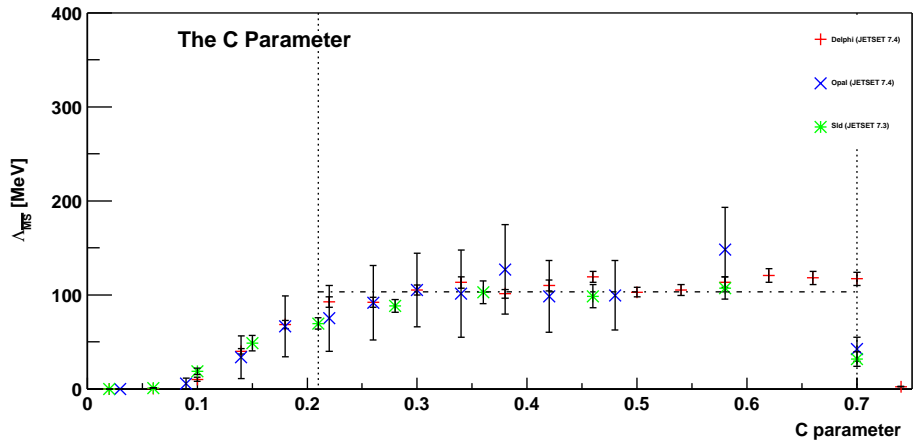,width=\textwidth}    
    \caption{The C Parameter using DELPHI, OPAL and SLD data}
    \label{fig:cpar.dlp}
  \end{center}
\end{figure}
\begin{figure}[!htbp]
  \begin{center}
    \epsfig{file=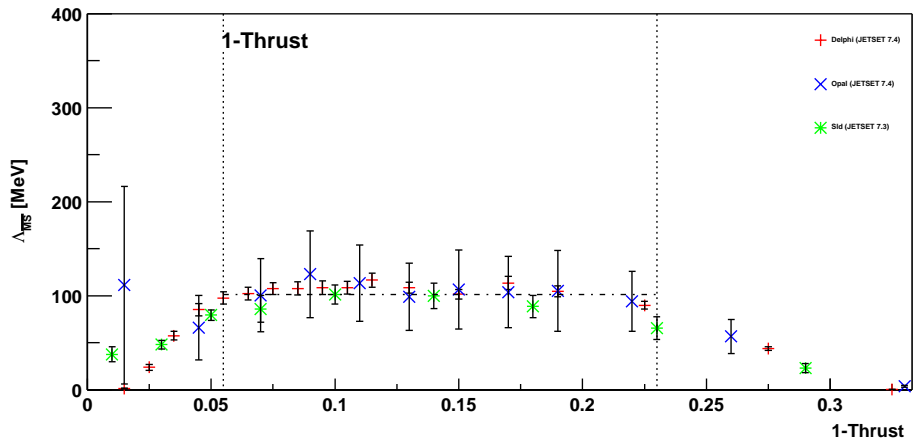,width=\textwidth}    
    \caption{The 1-Thrust Parameter using DELPHI, OPAL and SLD data}
    \label{fig:thrust.dlp}
  \end{center}
\end{figure}
\begin{figure}[!htbp]
  \begin{center}
    \epsfig{file=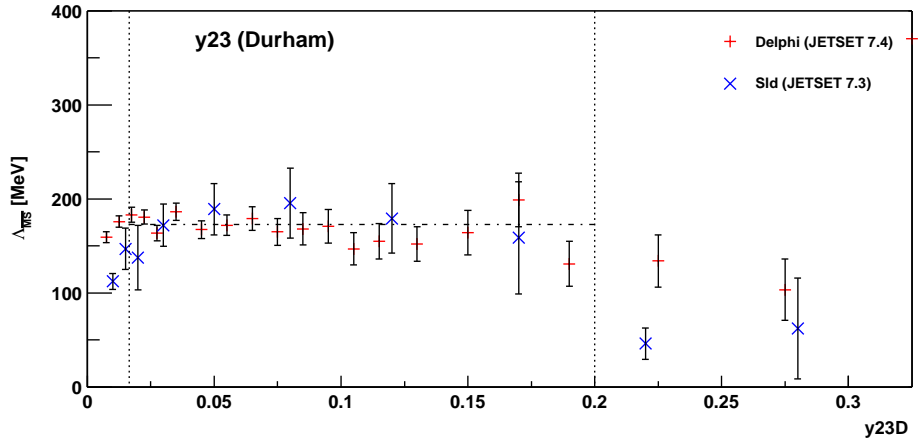,width=\textwidth}    
    \caption{The $2\to 3$ Jet Transition Parameter using DELPHI and SLD data}
    \label{fig:y23d.dlp}
  \end{center}
\end{figure}
\begin{figure}[!htbp]
  \begin{center}
    \epsfig{file=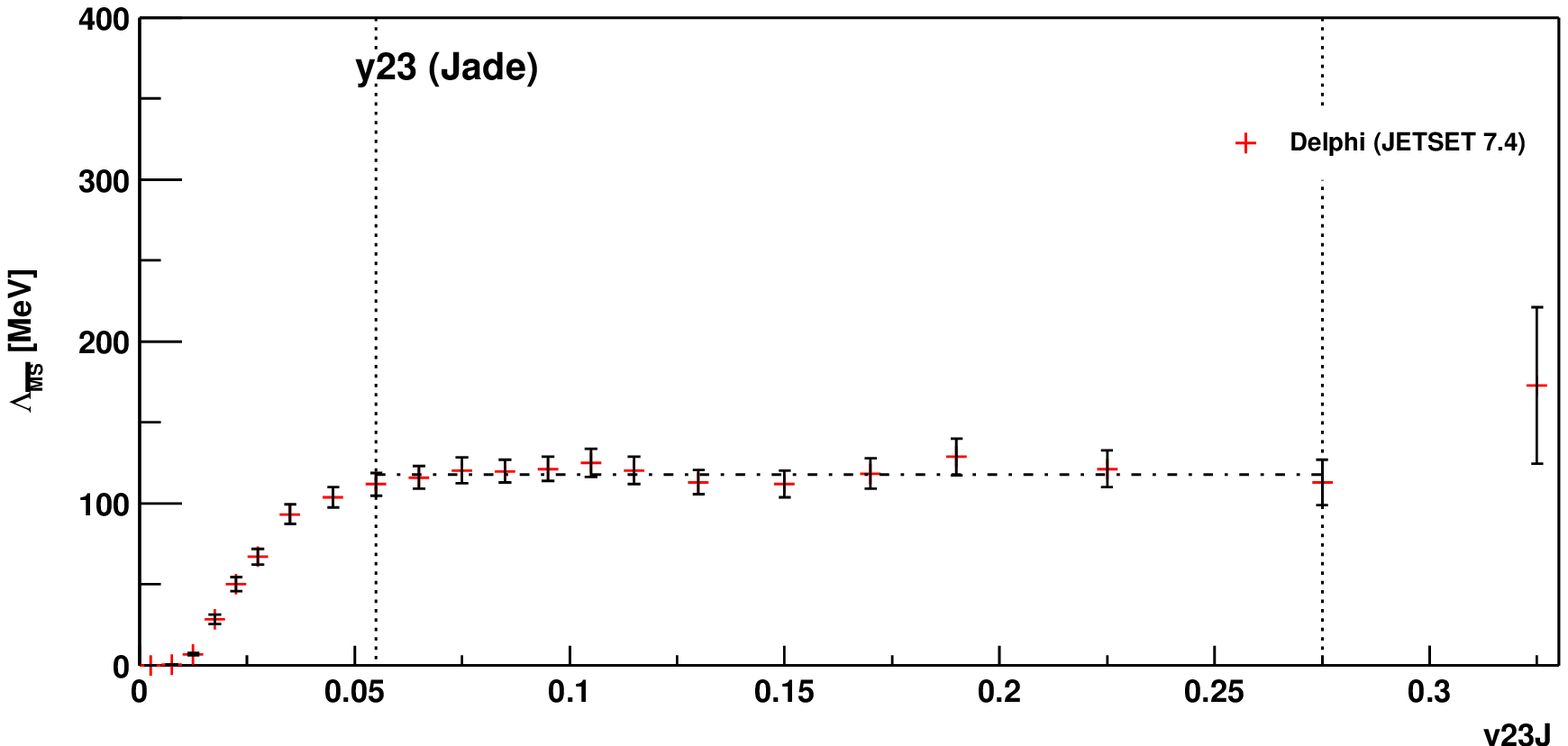,width=\textwidth}    
    \caption{The $2\to 3$ Jet Transition Parameter using DELPHI data}
     \label{fig:y23j.dlp}
  \end{center}
\end{figure}
\begin{figure}[!htbp]
  \begin{center}
    \epsfig{file=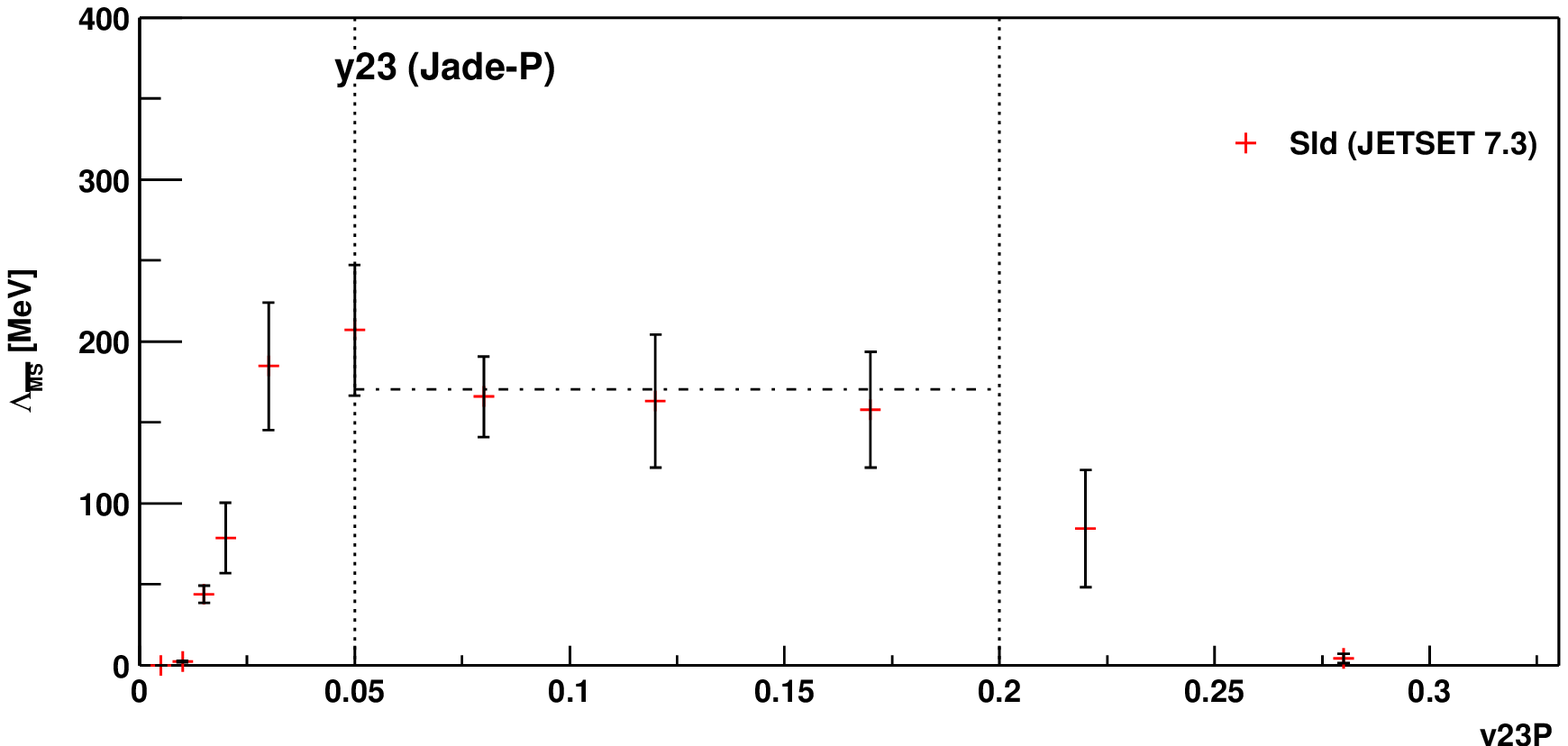,width=\textwidth}    
    \caption{The $2\to 3$ Jet Transition Parameter using SLD data}
    \label{fig:y23p.sld}
  \end{center}
\end{figure}
\begin{figure}[!htbp]
  \begin{center}
    \epsfig{file=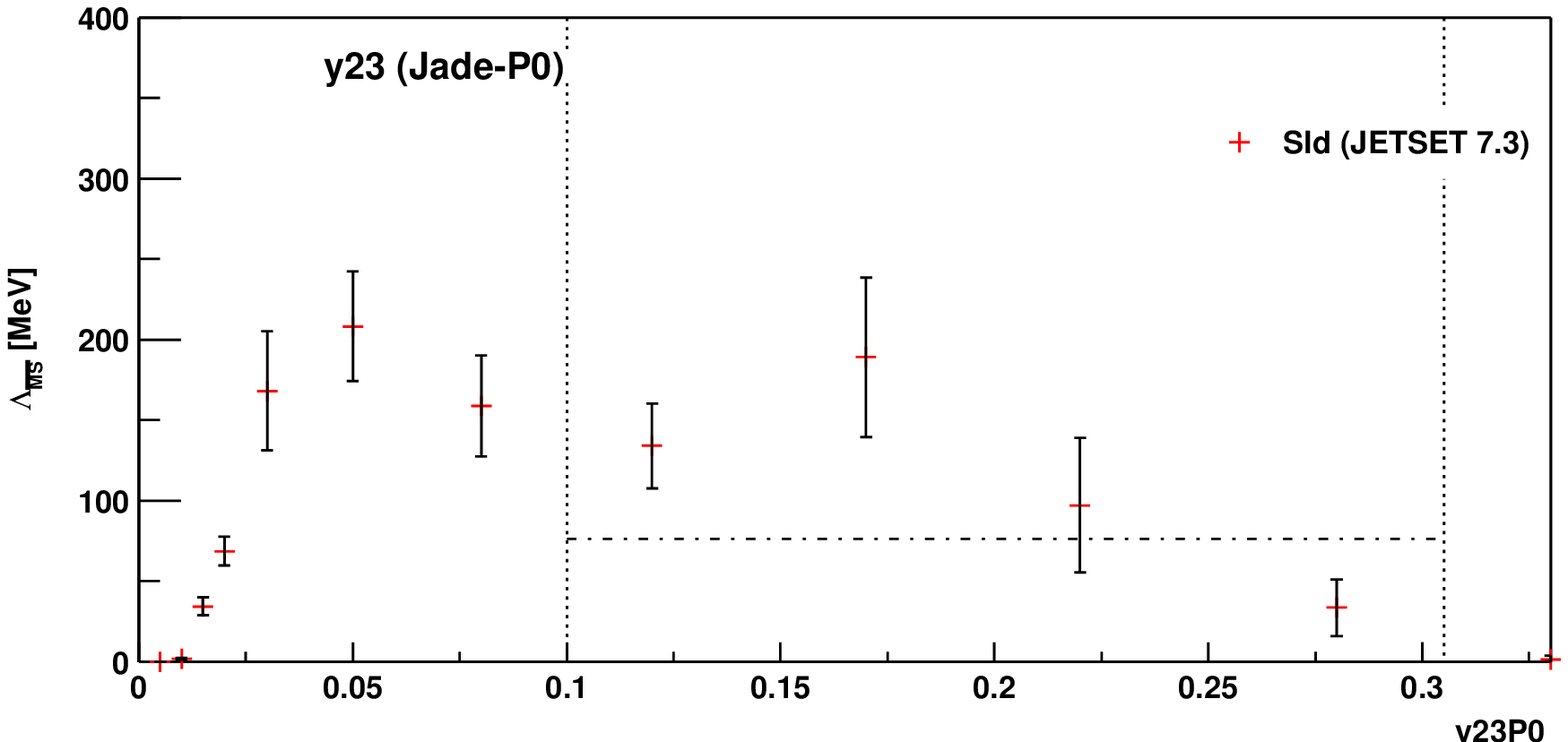,width=\textwidth}    
    \caption{The $2\to 3$ Jet Transition Parameter using SLD data}
    \label{fig:y23p0.sld}
  \end{center}
\end{figure}
\begin{figure}[!htbp]
  \begin{center}
    \epsfig{file=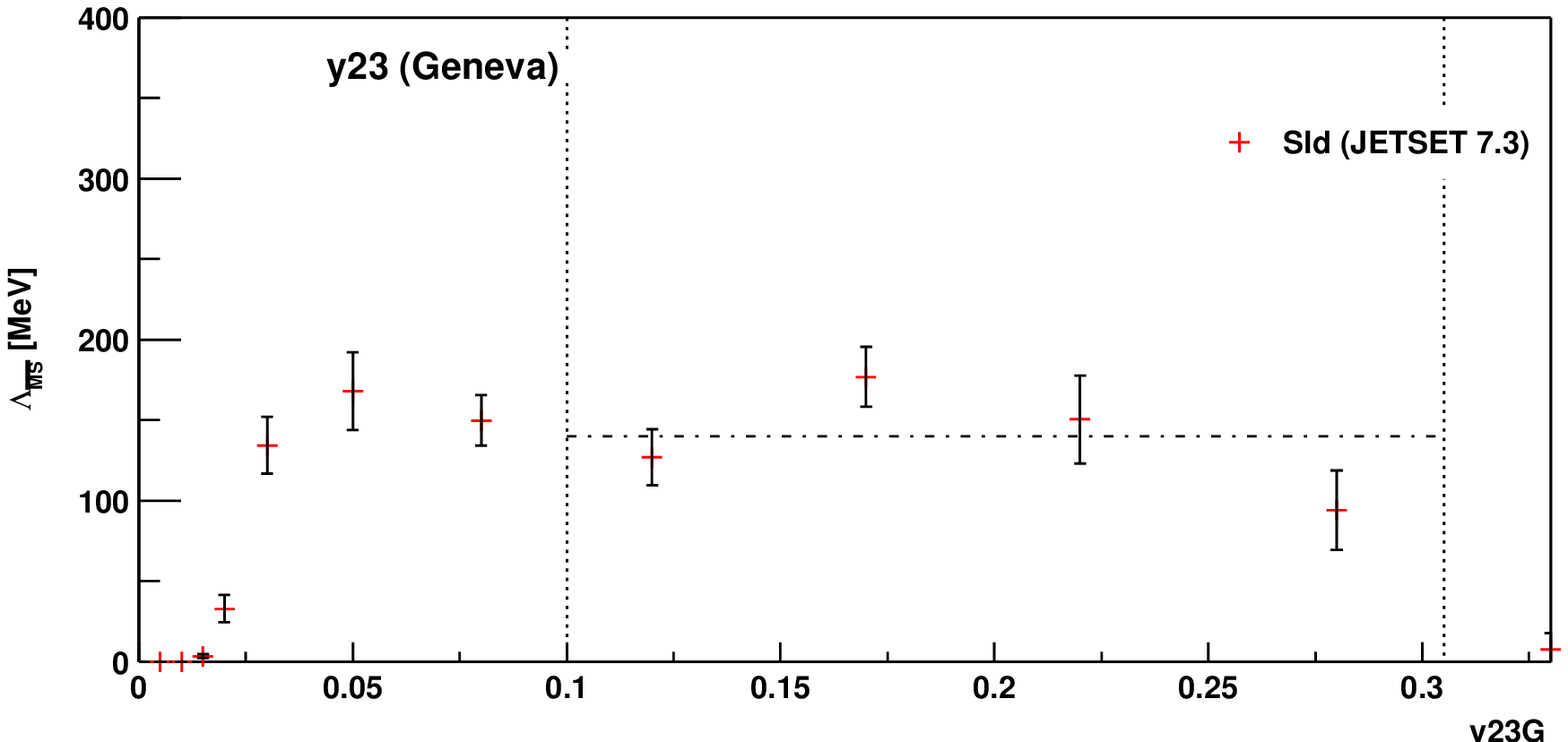,width=\textwidth}    
    \caption{The $2\to 3$ Jet Transition Parameter using SLD data}
    \label{fig:y23dg.sld}
  \end{center}
\end{figure}
\begin{figure}[!htbp]
  \begin{center}
    \epsfig{file=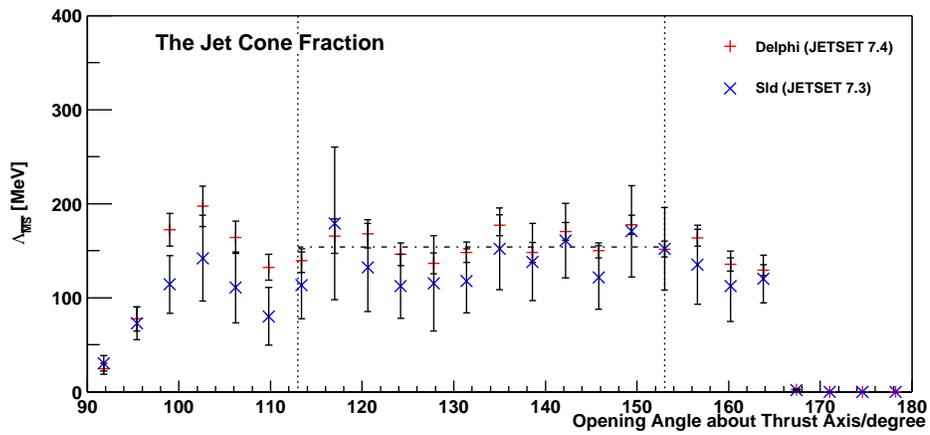,width=\textwidth}    
    \caption{The Jet Cone Energy Fraction using DELPHI and SLD data}
    \label{fig:jcef.dlp}
  \end{center}
\end{figure}
\vspace{-0.5cm}
\clearpage
We note that some of the variables described in Sec.~3, in particular the
difference in hemisphere masses, and the EEC and AEEC, have not been
included, since the fitting criteria outlined earlier are not satisfied.  For
comparison the corresponding fitted values using data {\it not} corrected for
hadronisation effects is given in Fig.~16.  Without hadronisation corrections
the fitted central values of ${\Lambda}_{\msb}$ lie in the range between 50
MeV and 275 MeV. Including hadronisation corrections, reduces this range to
between 100 MeV and 200 MeV. This factor of two uncertainty in
${\Lambda}_{\msb}$ is then to be taken as providing an estimate of the likely
size of remaining uncalculated NNLO and higher-order perturbative
corrections. If we assume that for {\it some} of the observables
${\cal{G}}({\cal{R}}){\approx}1$, then the factor of two uncertainty in
$\lmsb$ can be translated into an estimate of the potential size of the
two-loop NNLO RS-invariant ${\rho}_{2}$. Using Eq.(17) for
${\cal{G}}({\cal{R}})$ expanded as in Eq.(27), we see that
${\cal{G}}({\cal{R}}){\approx}2$, with the typical value of
${\cal{R}}{\approx}0.05$, corresponds to ${\rho}_{2}{\approx}50$.  Without
explicit NNLO calculations there is no rational basis for assigning a central
value for $\lmsb$ with an error. What one can say, however, is that inclusion
of hadronisation corrections does reduce the spread in the extracted $\lmsb$,
and that this remaining spread indicates a value of ${\rho}_{2}$ which is not
so large. We have no basis for estimating how big we might expect
${\rho}_{2}$ to be for these observables. We can note that for the $\epem$
$R$-ratio which has been computed to NNLO, ${\rho}_{2}=-15.1$ for ${N_f}=5$
active flavours \cite{r18}.
We can convert $\lmsb$ to ${\as}({M_Z})$ using the two-loop beta-function
equation. The corresponding values for hadronisation corrected data is given
in Fig.~18.
\begin{figure}[htbp]
  \begin{center}
    \epsfig{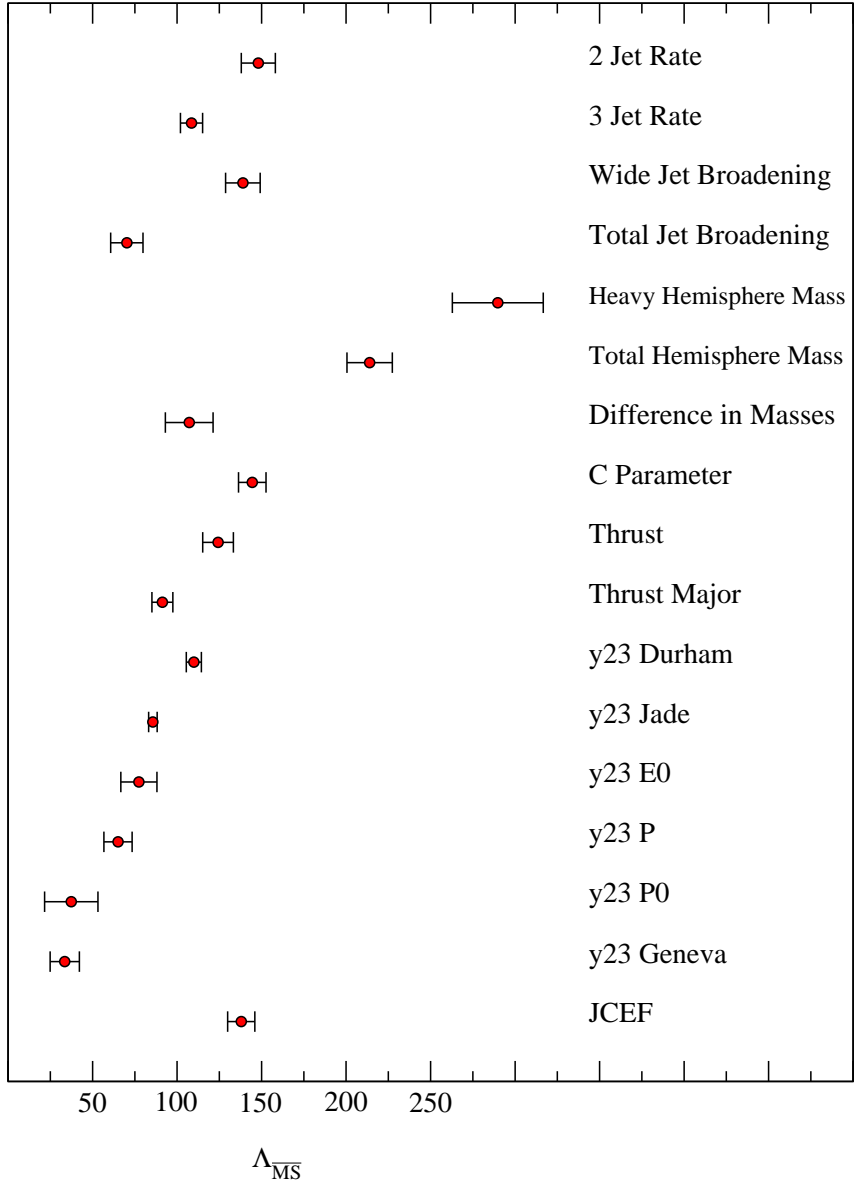}
    \caption{Summary of \lmsb\, measurements uncorrected for hadronisation effects.}
    \label{fig:3summ-raw}
  \end{center}
\end{figure}
\begin{figure}[htbp]
  \begin{center}
    \epsfig{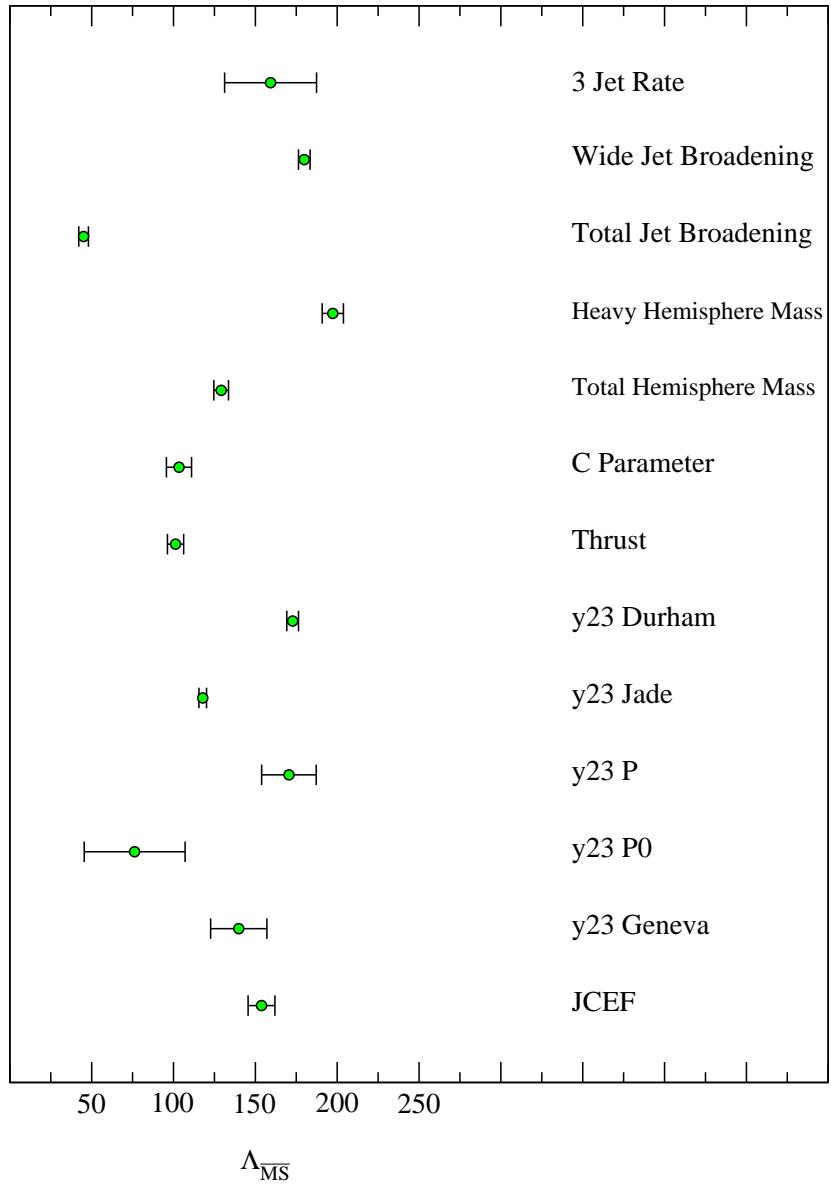}
    \caption{Summary of \lmsb\, measurements corrected for hadronisation effects.}
    \label{fig:3summ-hc}
  \end{center}
\end{figure}
\begin{figure}[htbp]
  \begin{center}
    \epsfig{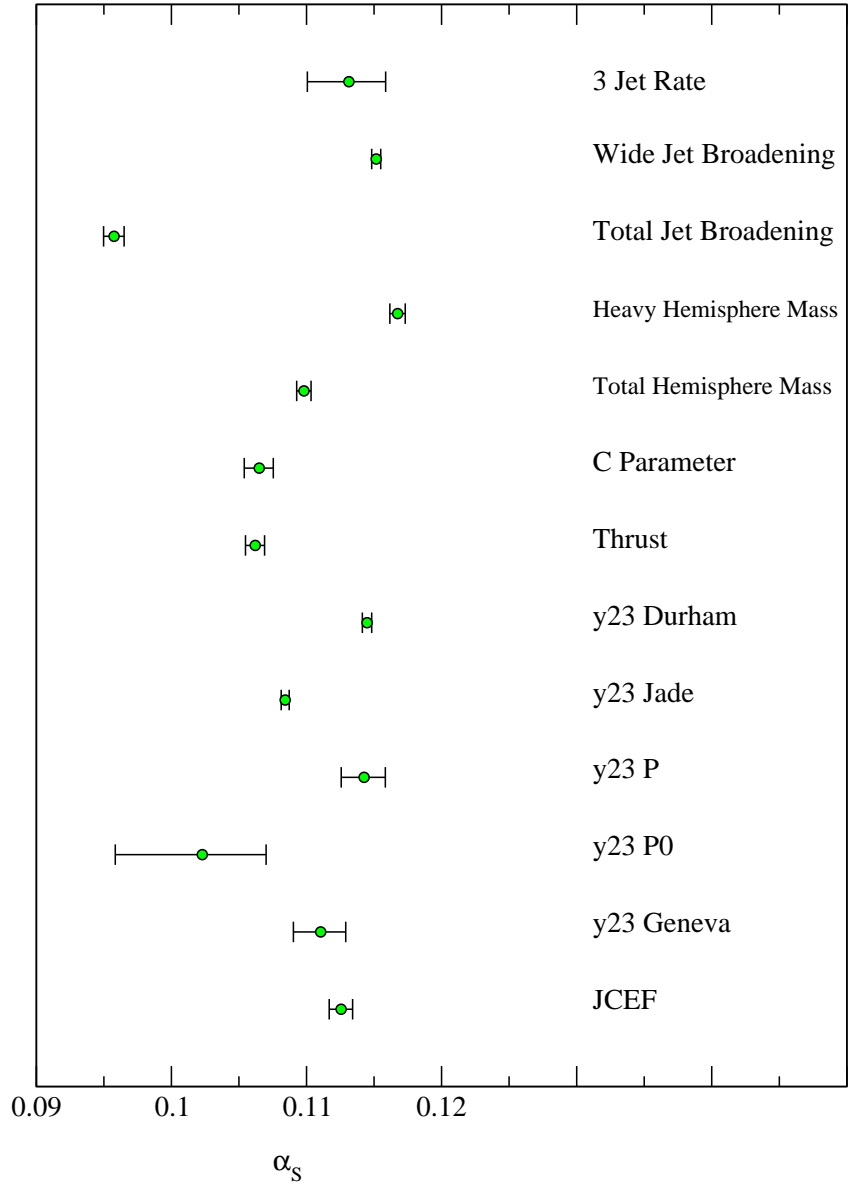}
    \caption{Summary of \as\ measurements corrected for hadronisation effects.}
    \label{fig:3alfsumm-hc}
  \end{center}
\end{figure}
\clearpage
We finally present the $\lmsb$ values obtained by fitting to the data on
four-jet observables. We use the DELPHI data of \cite{Abreu:1996na} in
all cases. In this case we only analysed data uncorrected for hadronisation
effects. The results for five variables are given in Fig.~19. With
the exception of the Light Hemisphere Mass and the Narrow Jet Broadening
the values are grouped between 150 and 200 MeV, reasonably consistent
with those obtained from the three-jet observables. It may be that
the Light Hemisphere Mass and Narrow Jet Broadening have rather
large hadronisation corrections.
\begin{figure}[hbp]
  \begin{center}
    \epsfig{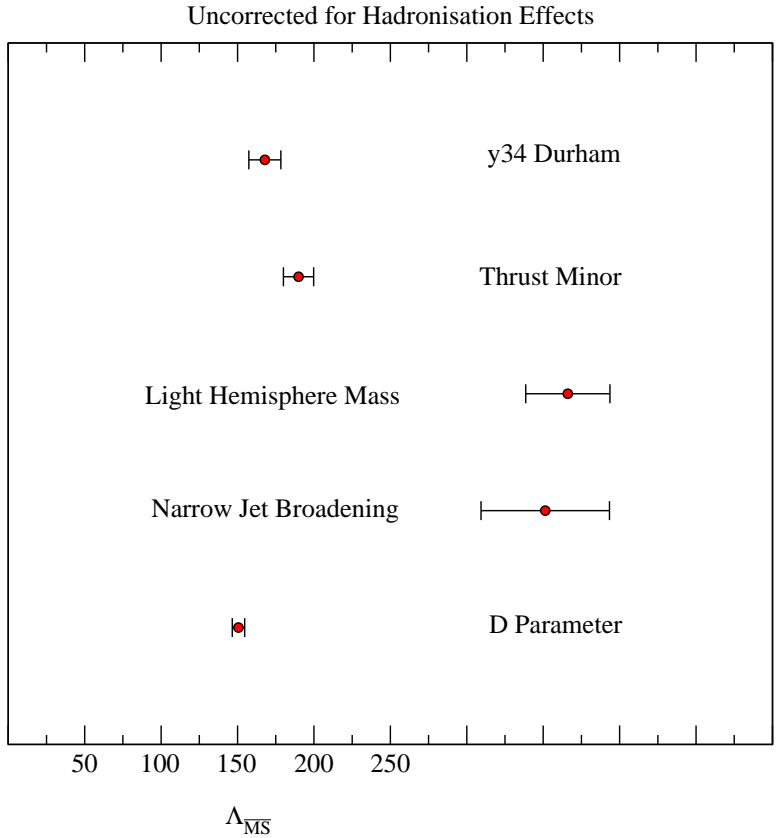}
    \caption{Summary of \lmsb\ \, measurements for four jet observables
      uncorrected for hadronisation effects.}
    \label{fig:4summ-raw}
  \end{center}
\end{figure}
\clearpage
\section*{5 Energy Dependence of the Thrust Distribution}
\label{sec:energy-depend-observ}
In this section we consider how to include power corrections in our
formalism. It is widely accepted that physical observables in general will be
subject to ``non-perturbative'' power-like corrections in the hard
interaction scale, $Q$. That is to say, there will be terms contributing to
cross-sections that cannot be expanded out in the typical perturbative manner
arising from expressions of the form
\begin{equation}
  \left(\frac{\Lambda}{Q}\right)^k = \exp\left(\frac{-k}{b\, \as}\right).
\end{equation}
Perturbative techniques cannot describe these terms accurately but have made
attempts at predicting the leading behaviour to the power corrections via
renormalon-inspired analysis \cite{Mueller:1992xz,Zakharov:1992bx} and
dispersive techniques \cite{Dokshitzer:1996qm}. Taking the generic form of
these power corrections, we can alter our perturbative expansion for a
dimensionless observable $\mR$,
\begin{equation}
  \label{eq:withpc}
  \mR = a + r_1 a^2 +r_2 a^3 + \ldots + \frac{\lambda_0}{Q} (1 + \lambda_1 a +
  \lambda_2 a^2 +\lambda_3 a^3 + \ldots ),
\end{equation}
where we have assumed a leading power correction with exponent 1 (i.e.
$\frac{1}{Q}$). To include these term in the analysis of Sec.~2, we must take the
derivative with respect to $Q$. We may then rewrite Eq.(8)
incorporating power corrections as 
\begin{eqnarray}
B({\mR}(Q))=\frac{\rd \mR}{\rd \ln Q} &=& -b \mR^2 (1+ c\mR + \rho_2\mR^2 + \rho_3\mR^3 +
\ldots)\nonumber\\
&+& \kappa_0 \mR^{-c/b} e^{-1/b \mR} ( 1+ \kappa_1 \mR + \ldots) + \ldots,  
\end{eqnarray}
where the $\kappa_n$ can be related to the $\lambda_n$. For example, the leading
power correction coefficient gives a $-\lambda_0/Q$ contribution to the
$\beta$-function. Using Eq.(12) to get $Q$ in the
leading approximation of $\mR$ we find,
\begin{equation}
Q \approx \Lambda_\mR e^{1/b\mR} (c \mR)^{c/b}.
\end{equation}
Substituting this back in we obtain 
\begin{equation}
  \lambda = - \kappa_0 e^{r_1(\mu=Q)/b} \left(\frac{b}{2}\right)^{c/b} \lmsb\,,
\end{equation}
where we have converted $\Lambda_R$ to $\lmsb\,$. Having made the connection
between $\kappa_0$ and $\lambda_0$ we may incorporate the power correction
term into Eq.(24) via the $\mathcal{G}(\mR(Q))$ function
given in Eq.(17). Expanding out $\mathcal{G}(\mR(Q))$ to
the accuracy of NNLO and leading power corrections gives
\begin{equation}
  \label{eq:Gpower}
  \mathcal{G}(\mR(Q)) \simeq 1 - \frac{\rho_2}{b}\mR +
  \frac{\kappa_0}{b}\frac{e^{-1/b\mR}}{\mR^2} \mR^{-c/b}.
\end{equation}
Substituting this back into Eq.(24) we finally obtain
\begin{equation}
  \label{eq:lampower}
  \lmsb = Q\,{\cal{F}}({\mR}(Q))\,\left(1 - \frac{\rho_2}{b}\mR +
  \frac{\kappa_0}{b}\frac{e^{-1/b\mR}}{\mR^2} \mR^{-c/b}\right) \,
  {e}^{-r/b}\, {\left(\frac{2c}{b}\right)}^{c/b}\;.
\end{equation}\\

\begin{figure}[t!]
  \begin{center}
    \epsfig{file=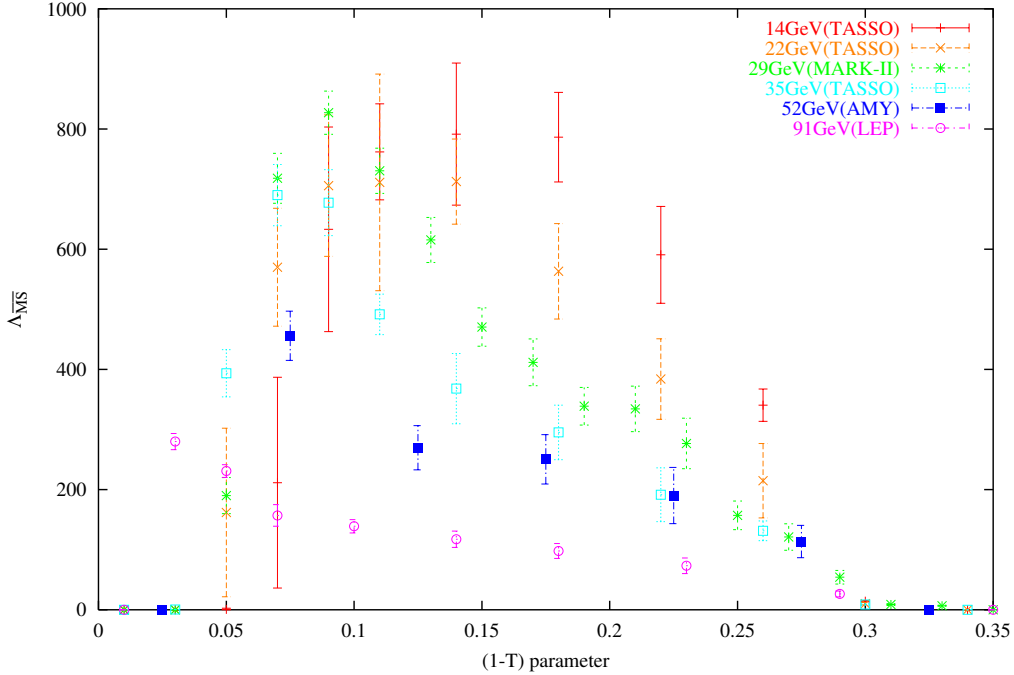,angle=270,width=\textwidth}
    \caption{The Energy Dependence of \lmsb\ Measurements from the Thrust Distribution}
    \label{fig:NRG-err}
  \end{center}
\end{figure}

\begin{figure}[t]
  \begin{center}
    \epsfig{file=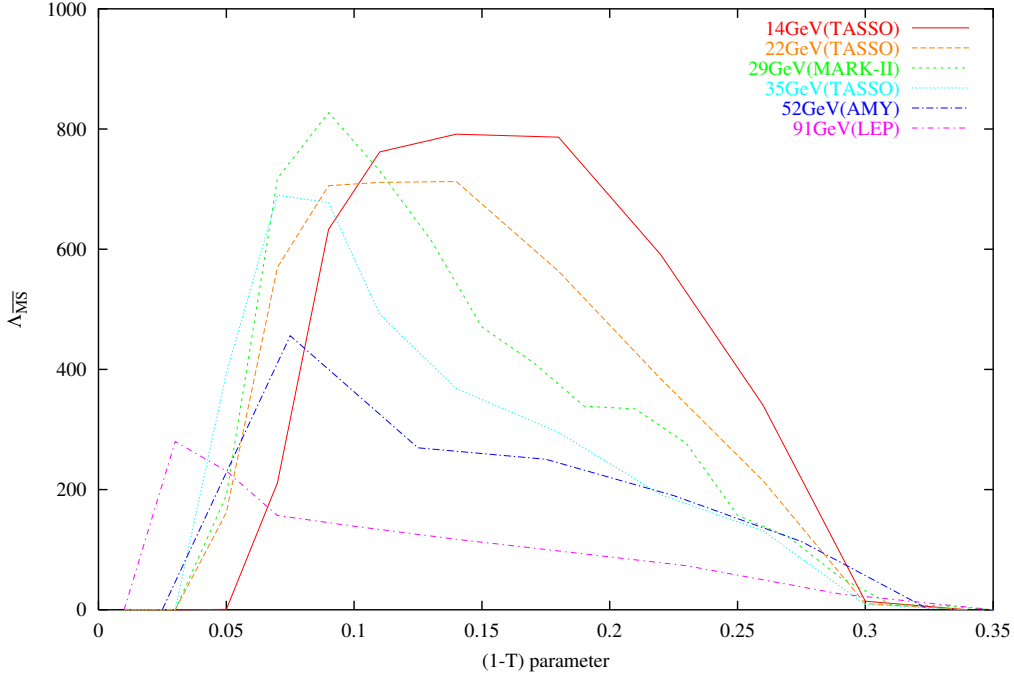,angle=270,width=\textwidth}
    \caption{The Energy Dependence of \lmsb\ Measurements from the Thrust Distribution}
    \label{fig:NRG-line}
  \end{center}
\end{figure}

We are fortunate enough to have experimental measurements of the thrust
distribution at a wide variety of energy scales from PETRA to LEP2.
We shall be considering the following data 
\begin{itemize}
\item PETRA (Detector- Tasso, Facility-DESY) \cite{Braunschweig:1990yd}
  \begin{itemize}
  \item Centre of Mass Energies - 14, 22, 35 GeV
  \end{itemize}
\item PEP (Detector- Mark-II, Facility-SLAC) \cite{Petersen:1988bq}
  \begin{itemize}
  \item Centre of Mass Energy - 29GeV
  \end{itemize}
\item TRISTAN (Detector- Amy, Facility-KEK) \cite{Li:1990sn}
  \begin{itemize}
  \item Centre of Mass Energy - 52GeV
  \end{itemize}
\item SLC (Detector- SLD, Facility-SLAC)
  \begin{itemize}
  \item Centre of Mass Energy - 91GeV
  \end{itemize}
\item LEP (Detectors- Aleph, Delphi, L3, Opal, Facility-CERN)
  \begin{itemize}
  \item Centre of Mass Energy - 91GeV
  \end{itemize}
\end{itemize}
Unfortunately as the LEP 2 data suffers from large errors due to poor
statistics, we are forced to exclude it from the analysis.  
We would expect that the ${\lmsb}$ directly extracted from Eq.(25)
should approach the actual value as the energy increases, and sub-asymptotic
effects become smaller with ${\mG}(\mR)$ approaching more closely its
asymptotic value of unity. This trend can indeed be seen from
Figs.~\ref{fig:NRG-err} and~\ref{fig:NRG-line} (where we have removed the
error bars for clarity). In the limit $1-T\to \frac{1}{3}$ we are faced once
again with the problem of the kinematic end point dragging the value to zero.
In the $1-T\to 0$ limit, large kinematic logarithms dominate from the
emission of soft and collinear gluons. Clearly in between we see a
``flattening'' of the \lmsb\, value with $1-T$. This can be interpreted as
higher order and power-like corrections having less influence, and hence the
NLO approximation becoming more reliable. 
\\

\begin{figure}[tbp]
  \begin{center}
    \epsfig{file=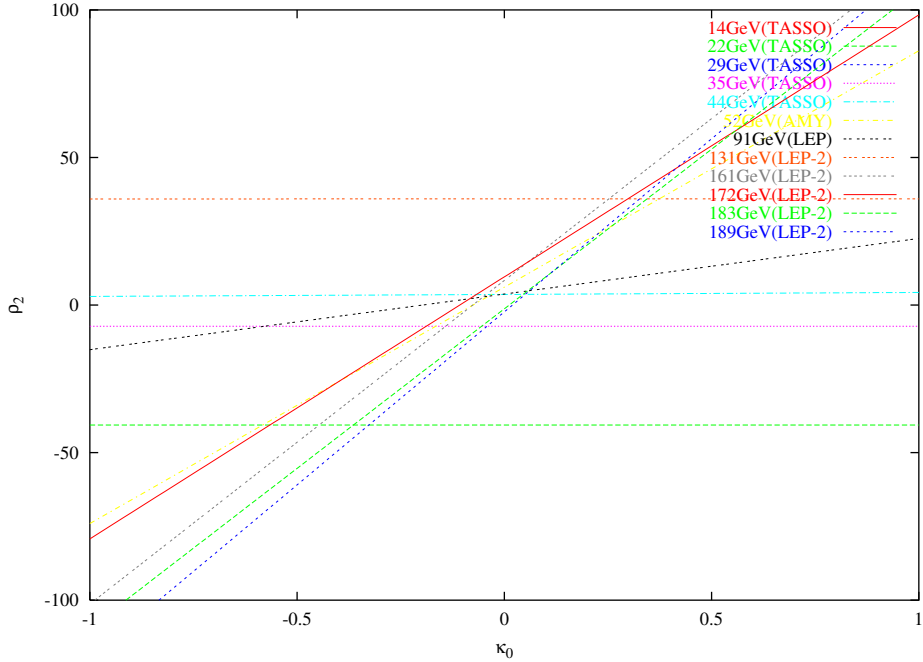}
    \caption{Fitting for the leading power correction ($\kappa_0$) and NNLO
      ($\rho_2$) using the thrust distribution at different energies. We use
      a value of \lmsb\ = 200MeV } 
    \label{fig:thrust-power}
  \end{center}
\end{figure}

It is possible to consider the formalism from a different perspective whereby
we accept that \lmsb\, is a constant and that any deviations from its true
value are due to higher order effects and power corrections. In this case we
may then try to approximate these contributions by the leading terms as given
by Eq.~(\ref{eq:lampower}). In this section we highlight a simple mechanism
for investigating these effects.  We stress that this by no means provides a
highly accurate estimate of these terms, but merely an indication of how
important they are. We begin by rewriting Eq.~(\ref{eq:lampower}) as
\begin{equation}
  \kappa_0 = M(\mR,r) \rho_2 +C(\mR,r),
\end{equation}
where $M(\mR,r)$ and $C(\mR,r)$ are known terms at NLO, having specified a
value for \lmsb\,. This is simply the equation of a straight line in
$(\kappa_0,\rho_2)$ space. Since these two quantities are $Q$-independent, we
may plot the lines corresponding to different centre-of-mass energies and
expect them to cross over at the solution. This is illustrated in
Fig.~\ref{fig:thrust-power} where we have taken a value of
$\lmsb\,=200$MeV. We have made no attempt to incorporate the errors, which
for the LEP 2 data will be considerable. In all cases, the central value is
taken. This na{\"\i}ve procedure does give a promising result, though. There
appear to be two predominant localised cross-over regions where the lines
appear to converge. These correspond to ${\rho}_{2}$ and ${\kappa}_{0}$ both
small or ${\rho}_{2}{\approx}{-40}$ and ${\kappa}_{0}{\approx}-0.5$. Similar
fits for the $<1-T>$ were performed in \cite{camglovmax}.

\section*{6  Discussion and Conclusions}

In this paper we have shown that one can directly relate QCD observables to
the underlying dimensional transmutation parameter of the theory, leading to
the relation of Eq.(24). This derivation does not require the specification
of a renormalisation scale, or use of the convention-dependent renormalised
coupling ${\as}(\mu)$. It is obtained simply by integrating up the obvious
dimensional analysis statement of Eq.(3). This integration corresponds to
identifying and resumming the complete set of ultra-violet logarithms of $Q$
which reside in the perturbative coefficients. In standard RG-improvement an
infinite subset of these is tacitly omitted, and this results in the
problematic scale-dependence of fixed-order predictions which plagues
standard analyses. The only approximation in the present approach is the
incomplete knowledge of the function ${\mG}(\mR)$ which approaches unity for
asymptotic $Q$, and deviates from it by two-loop (NNLO) sub-asymptotic
effects characterised by the NNLO RS-invariant ${\rho}_{2}$, and by power
corrections. Given that these are presently unknown for the QCD $\epem$ jet
observables we wished to analyse, the best we could do was to use Eq.(25),
corresponding to the asymptotic expectation ${\mG}({\mR}(Q)){\approx}1$. As
we discussed in Sec.~4, to avoid regions containing large kinematical
logarithms arising in the two-jet limit, and kinematical endpoints, it was
necessary to choose the range in the observable over which the fits to
constant $\lmsb$ were performed, rather carefully. Notwithstanding the fact
that these fitting criteria are inevitably somewhat {\it ad hoc}, the
procedure crucially avoids the additional scatter in extracted ${\lmsb}$
values resulting from the arbitrary variation of the renormalization scale,
which is customarily employed. One finds that the range of directly fitted
$\lmsb$ values obtained is significantly reduced using hadronisation
corrected data (see Figs.~16 and~17). The remaining scatter in $\lmsb$
values displayed in Fig.~17 is then to be attributed to the presence of
uncalculated NNLO and higher-order sub-asymptotic effects in ${\mG}({\mR})$.
The factor of two scatter in $\lmsb$, roughly between $100$ and $200$ MeV,
then corresponds on using the exponentiated form for ${\mG}({\mR})$ in
Eq.(27), to an estimate $|{\rho}_{2}|{\approx}50$, indicating significant but
not huge uncalculated two-loop effects. A firmer statement about the value of
$\lmsb$ is only possible once these effects have been computed. There seems
to us to be no point in extracting a purported central value of $\lmsb$ together with
an error in the meantime.\\

 In Sec.~5 we performed a complementary exercise
in which, assuming a fixed value of ${\lmsb}=200$MeV, the $Q$-dependence
of the thrust distribution over the energy range from PETRA to LEP 2 was
used to fit for the NNLO RS-invariant ${\rho}_{2}$ and power corrections.
As seen from Fig.~21 as $Q$ increases the directly extracted $\lmsb$
is reduced, and its distribution in thrust flattened, as would be
expected if the sub-asymptotic effects contained in ${\mG}({\mR})$
are becoming smaller. The fits for ${\rho}_{2}$ and the ${\kappa}_{0}$,
the parameter controlling $1/Q$ power corrections, shown in Fig.~22
reveal two possibilities. One corresponds to both ${\rho}_{2}$ and
${\kappa}_{0}$ small, and the other to ${\kappa}_{0}$ small and
${\rho}_{2}{\approx}-50$, a value consistent with that estimated
from the scatter of $\lmsb$ values in Fig.~17.\\

There are two major directions in which the formalism described
here could be improved. The first would be to resum to all-orders
large kinematical infra-red logarithms in the function ${\mG}({\mR})$. This
would allow fitting over a much larger range in the observables.
Whilst in principle straightforward a number of technical complications
present themselves. The second would be to generalize the direct
extraction of ${\lmsb}$ to see how Eq.(24) is modified if one has
massive quarks. We hope to report developments in both these areas
in future papers.

\newpage
\section*{Acknowledgements}
We would like to thank Nigel Glover for crucial help in generating the NLO
perturbative corrections for the observables considered from the Monte Carlo
program of \cite{Giele:1992vf,Campbell:1998nn}, and for numerous interesting
and stimulating discussions on the CORGI approach. We are grateful to Phil
Burrows (SLD), Siggi Hahn (DELPHI) and Otmar Biebel (OPAL) for generously
providing hadronisation corrections for their experimental collaboration's
respective data sets. This work was supported in part by the EU Fourth
Framework Programme `Training and Mobility of Researchers', Network `Quantum
Chromodynamics and the Deep Structure of Elementary Particles', contract
FMRX-CT98-0194 (DG-12-MIHT).

\end{document}